\PassOptionsToPackage{dvipsnames}{xcolor}
\documentclass[epjc3,twocolumn]{svjour3}

\usepackage{amsmath,amsfonts,booktabs,cite,graphicx,listings,orcidlink,relsize,slashed,url,xspace}
\usepackage[utf8]{inputenc}

\usepackage{tabularx}

\definecolor{lightgrey}{gray}{0.9}

\def\eg{{\it e.g.}}
\def\ie{{\it i.e.}}

\usepackage{siunitx}
\newcommand{\electronvolt}{\text{e\kern-0.15ex V}}
\DeclareSIUnit{\eV}{\electronvolt}
\DeclareSIUnit{\MeV}{M\electronvolt}
\DeclareSIUnit{\GeV}{G\electronvolt}
\DeclareSIUnit{\TeV}{T\electronvolt}
\DeclareSIUnit\barn{b}
\DeclareSIUnit{\ifb}{\femto\barn\tothe{-1}}
\DeclareSIUnit{\iab}{\atto\barn\tothe{-1}}

\newcommand{\ma}{\textsc{MadAnalysis~5}\xspace}
\newcommand{\madgraph}{\textsc{MadGraph5\_aMC@NLO}\xspace}

\newcommand{\rivet}{\textsc{Rivet}\xspace}

\newcommand{\htt}{\textsc{HEPTopTagger}\xspace}

\newcommand{\pythia}{\textsc{Pythia}\xspace}

\newcommand{\fastjet}{\textsc{FastJet}\xspace}
\newcommand{\fjcontrib}{\textsc{FastJet Contrib}\xspace}

\newcommand{\pt}{\ensuremath{p_\mathrm{T}}\xspace}
\newcommand{\tL}{\ensuremath{t_\mathrm{L}^\text{rec}}\xspace}
\newcommand{\tH}{\ensuremath{t_\mathrm{H}}\xspace}
\newcommand{\WL}{\ensuremath{W_\mathrm{L}^\text{rec}}\xspace}
\newcommand{\bL}{\ensuremath{b_\mathrm{L}}\xspace}

\newcommand{\ttbar}{{\ensuremath{t\bar{t}}}\xspace}
\newcommand{\akt}{\text{anti-$k_T$}\xspace}

\smartqed  
\journalname{Eur. Phys. J. C}

\hypersetup{
  unicode=true,pdftoolbar=false,pdfmenubar=true,
  pdffitwindow=false,pdfnewwindow=true,
  pdfstartview={FitH},
  pdfauthor={J.~Araz, A.~Buckley, B.~Fuks},
  pdfkeywords={SMEFT LHC top-quark boost tagging reconstruction HEPTopTagger Rivet MadAnalysis},
  pdftitle={Searches for new physics with boosted top quarks in the MadAnalysis5 and Rivet frameworks},
  colorlinks=true,linkcolor=Maroon,citecolor=Maroon,filecolor=Maroon,urlcolor=ForestGreen
}

\begin{document}
\title{Searches for new physics with boosted top quarks in the \ma and \rivet frameworks}

\author{
  Jack~Y.~Araz\thanksref{e1,addr1}\orcidlink{0000-0001-8721-8042}
  \and\
  Andy~Buckley\thanksref{e2,addr2}\orcidlink{0000-0001-8355-9237}
  \and\
  Benjamin~Fuks\thanksref{e3,addr3}\orcidlink{0000-0002-0041-0566}
}

 \thankstext{e1}{E-mail: {\color{RedViolet}jack.araz@durham.ac.uk}}
 \thankstext{e2}{E-mail: {\color{RedViolet}andy.buckley@glasgow.ac.uk}}
 \thankstext{e3}{E-mail: {\color{RedViolet}fuks@lpthe.jussieu.fr}}

 \institute{
  Institute for Particle Physics Phenomenology, Durham University, Durham, UK\label{addr1}
  \and
  School of Physics and Astronomy, University of Glasgow, Glasgow, UK\label{addr2}
  \and\
  Laboratoire de Physique Th\'eorique et Hautes \'Energies (LPTHE), UMR 7589, Sorbonne Universit\'e et CNRS, 4 place Jussieu, 75252 Paris Cedex 05, France \label{addr3}
}

\date{Received: date / Accepted: date}

\maketitle

\begin{abstract}
  High-momentum top quarks are a natural physical system in collider experiments for testing models of new physics, and jet substructure methods are key both to exploiting their largest decay mode and to assuaging resolution difficulties as the boosted system becomes increasingly collimated in the detector. To be  used in new-physics interpretation studies, it is crucial that related methods get implemented in analysis frameworks allowing for the reinterpretation of the results of the LHC such as \ma and \rivet. We describe the implementation of the \htt algorithm in these two frameworks, and we exemplify the usage of the resulting functionalities to explore the sensitivity of boosted top reconstruction performance to new physics contributions from the Standard Model Effective Field Theory. The results of this study lead to important conclusions about the implicit assumption of Standard-Model-like top-quark decays in associated collider analyses, and for the prospects to constrain the Standard Model Effective Field Theory via kinematic observables built from boosted semi-leptonic $t\bar{t}$ events selected using \htt.
\end{abstract}


\section{Introduction}\label{sec:intro}

Since the resurrection of jet-substructure methods as probes for new particles at the LHC~\cite{Seymour:1993mx, Butterworth:2008iy}, boosted topologies in which multiple decay products from heavy intermediate states fall into a single large-radius (large-$R$) jet have seen wide application in searches for new physics~\cite{Kaplan:2008ie, Ellis:2009su, Abdesselam:2010pt, Larkoski:2017jix, Kogler:2018hem, Marzani:2019hun}. While not initially considered in the early days of the LHC, these jet substructure techniques are now indeed largely used to extend the sensitivity of searches for new physics. This is particularly the case as the currently null results of those searches indicate that any relevant physics beyond the Standard Model (BSM) is most-likely located at a large mass scale, featuring heavy particles whose production and decay would naturally yield highly-boosted lighter Standard Model (SM) objects.

Many collider signatures can benefit from the usage of jet substructure methods, as they can be generally applied to tag many SM and BSM particles when they are produced with a high Lorentz boost. Among these, the top quark is an important target, for two reasons. First, the top quark is the highest-mass fermion in the SM, featuring a Yukawa coupling value close to~1. This makes it a natural candidate to provide an explanation for the hierarchy problem, and to play the role of a mediator that couples to new-physics sectors (\eg~through the Higgs field). Second, boosted methods can provide better background-rejection power than a classic `resolved' reconstruction of the top-quark kinematics. As a high-mass, colour-charged and non-hadronising particle, the top quark is the most complex SM resonance to reconstruct from fully resolved decay components. This not only requires highly performant $b$-tagging, but also suffers from either a complicated lepton and missing-momentum reconstruction or the resolution difficulties inherent to reconstructing a fully hadronic $b\bar{q}q'$ final state.

Jet-substructure methods offer a way to bypass many of the difficulties related to the reconstruction and identification of hadronically-decaying top quarks by relying on one single large-radius jet in place of three small-radius ones. In addition, such an option generally exploits the presence of \emph{two} heavy SM particles' decay hierarchies within the large-$R$ jet (the top quark itself and the $W$ boson originating from its decay), together with information on the internal momentum and angular structure of all jet constituents (with or without $b$-tagging requirements) to disambiguate boosted top quarks from jets originating from pure QCD background processes. A prominent tool in such studies is the \htt method~\cite{Plehn:2010st,Kasieczka:2015jma}, which pioneered this approach and has since gone through several rounds of enhancement such as use of variable-radius jet clustering. In the meantime more sophisticated and efficient top tagging methods have been developed. Typical examples are based on a classification of jets making use of the radiation pattern within a jet (also known as shower deconstruction)~\cite{Soper:2012pb}, on advanced machine learning techniques (we refer to Ref.~\cite{Kasieczka:2019dbj} for an overview) relying on observables like the jet transverse momentum and mass, the dispersion of its constituents estimated through $N$-subjettiness variables~\cite{Thaler:2010tr, Thaler:2011gf}, splitting scales~\cite{Thaler:2008ju}, energy correlation functions~\cite{Larkoski:2013eya, Larkoski:2014gra}, as well as on jet image analysis by means of neutral networks~\cite{Cogan:2014oua, deOliveira:2015xxd, Baldi:2016fql} and image or language recognition techniques~\cite{Almeida:2015jua, Kasieczka:2017nvn, Macaluso:2018tck, Araz:2021wqm}. More recently, a series of machine-learning methods embedding Lorentz invariance~\cite{Gong:2022lye, Bogatskiy:2022czk} have additionally been proposed and explored. The \htt method, however, still plays the role of being an important benchmark in the top-tagging landscape, especially in the context of use by the LHC experiments~\cite{ATLAS:2018wis, CMS:2020poo}. On the other hand, the related code has historically been unavailable for use in analysis prototyping and preservation within the two public analysis frameworks \ma~\cite{Conte:2012fm, Conte:2014zja, Conte:2018vmg} and \rivet~\cite{Buckley:2010ar, Bierlich:2019rhm}, that are widely used across the high-energy physics community. The goal of the present work is to fill this gap, and to document, through a few examples, its addition to both frameworks. It also therefore serves as a prototype interface for integration of C++ versions of machine-learning taggers into these public analysis toolkits.

While most applications of boosted top-quark reconstruction have been aimed at direct searches for new physics, the lack of tangible evidence for new high-mass resonances urges complementary studies of indirect routes through which BSM physics can manifest. A leading approach in this is that of the Standard Model Effective Field Theory (SMEFT), in which the explicit microscopic physics of a particular BSM model is replaced by an infinite set of higher-dimensional operators involving the SM fields and compatible with the SM symmetries~\cite{Weinberg:1978kz, Leung:1984ni, Buchmuller:1985jz}. The SMEFT is then an expansion in an energy scale $\Lambda$ above which the effective theory breaks down and real new physics resides, so that new fields with masses comparable with $\Lambda$ must be explicitly added to the model's Lagrangian. Details about the UV theory are encoded in the Wilson coefficients multiplying each operator, and the relevance of a specific new interaction is dictated ({\it a priori}) by the dimension of the corresponding operator (that is thus suppressed by some power of the effective scale). At dimension six, 84 (3045) parameters encode the leading BSM effects, assuming a flavour-blind (flavour-general) setup~\cite{Grzadkowski:2010es, Henning:2015alf}. Constraints are typically made primarily in the model-independent space of the corresponding Wilson coefficients by investigating the possibility of small and often subtle deviations from the SM expectation.  Among all operators, about twenty of them impact top physics under the simplifying assumption that new physics couples dominantly to bosons and to the left-handed doublet and right-handed up-type singlet of third generation quarks~\cite{AguilarSaavedra:2018nen}.

Global SMEFT interpretations of measurements at the LHC in the top sector have recently been achieved by several groups~\cite{Buckley:2015lku, Hartland:2019bjb, Brivio:2019ius, Ellis:2020unq, Ethier:2021bye,Giani:2023gfq}. These studies demonstrated in particular that dozens of SMEFT operators could be constrained (and therefore determined) simultaneously, correlating sometimes information originating from different sectors. It is nevertheless well known that signatures of processes involving boosted top quarks could be crucially relevant~\cite{Englert:2016aei}. These indeed involve large momentum transfers, so that they are expected to exhibit the largest sensitivity to new physics effects in the SMEFT, and subsequently show the most sensitivity to BSM phenomena.  It is therefore natural to focus on
high-momentum collider event categories involving the production of boosted top quarks, and to consider them as a promising avenue to statistically constrain the viable space of Wilson coefficients associated with top quark operators.

In the present study, we make use of the \htt functionalities that we implemented in the \rivet and \ma frameworks (together the possibility of computing emulated reconstruction-level observables) to study the sensitivity of the LHC to top-related SMEFT operators, focusing on the production of a pair of boosted top quarks. However, the \htt algorithm is designed to exploit as much as possible the kinematics of the SM decay of a boosted top quark. This leads to the open question about how new physics effects arising from the introduction of non-zero top-quark SMEFT operators could modify these kinematics, and hence impact the performance of \htt and, by inference, of any similar reconstruction method based on the topology arising from an SM top quark decay. As a straightforward application and keeping this in mind, we highlight important resulting issues for BSM interpretations.

In Section~\ref{sec:httanacode}, we detail our technical developments in \rivet (Section~\ref{sec:rivet}) and \ma (Section~\ref{sec:ma5}), and briefly explain how to use the codes for physics studies. In Section~\ref{sec:top_eft}, we exemplify the usage of these developments to estimate the impact of new physics via effective SMEFT operators on \htt performance, and how this affects the sensitivity of the present and future runs of the LHC (assuming an integrated luminosity of \SI{300}{\ifb} and \SI{3000}{\ifb} and varied levels of systematic errors) to these operators. We summarise our work in Section~\ref{sec:conclusion}.

\section{\htt implementation in the \rivet and \ma frameworks}\label{sec:httanacode}

\subsection{Generalities}
In its initial proposal~\cite{Plehn:2010st}, the \htt algorithm is a purely deterministic top-tagging method in which boosted top reconstruction is solely achieved from the geometrical structure and properties of the constituents of a fat jet. It first defines a fat jet collection from an event final state by using the Cambridge-Aachen jet algorithm~\cite{Dokshitzer:1997in, Bentvelsen:1998ug, Wobisch:1998wt}. A procedure is next applied to all jets included in this collection, in order to decide whether they should be top-tagged.

In practice, each reconstructed fat jet is decomposed into several subjets by applying a mass drop criterion~\cite{Dasgupta:2013ihk}. More precisely, jet clustering is iteratively undone so that each fat jet is split in two subjets, and the subjet with the smallest invariant mass is kept only if its invariant mass is large enough. Each resulting subjet is further decomposed in the same manner provided that its invariant mass is larger than some threshold. All possible triplets of jets belonging to the subjet collection obtained in this way are then filtered, and the five hardest filtered subjets are selected for boosted top quark reconstruction.

These five subjets are reclustered into three subjets, that are thus assumed to originate from a top quark decay. Events are at this stage rejected if they do not include any resulting triplet with an invariant mass that is compatible with the top mass. Top tagging stems from several requirements that are imposed on the invariant masses of the different dijet pairs that could be formed from the three subjects of any boosted top quark candidate, in particular in order to ensure the compatibility with the presence of an intermediate $W$ boson. Moreover the transverse momentum of the top candidate is required to be at least 200~GeV. 

We refer to the original documentation~\cite{Plehn:2010st, Kasieczka:2015jma} for a more comprehensive and quantitative presentation of the \htt algorithm.

The performance of any top-quark tagger can be improved by using an increased set of input variables (as in most multi-variate methods), for which the explicit choices are made through a tuning process relative to a given reference. To this end, the \htt method has been updated and now includes a variety of features that enhance the tagging efficiency and reduce the associated mistagging rates: it uses substructure mass-drop conditions~\cite{Dasgupta:2013ihk}, jet trimming~\cite{Krohn:2009th} and pruning~\cite{Ellis:2009su,Ellis:2009me} algorithms, and filtering steps~\cite{Butterworth:2008iy}, in addition to the core requirement that the large-radius jet demonstrates the three-pronged structure characteristic of a boosted top-quark's hadronic decay. In the current version of the \htt package, all these methods are used together in a multi-variate classification~\cite{Chien:2013kca,Ellis:2014eya} which maximises the expected tagging performance.

Access to this tool and all its embedded features within public frameworks like \ma or \rivet is thus crucial for prototyping and reproducing collider-event data analyses, a key activity in collider phenomenology. In the rest of this section, we discuss technical details about the embedding of \htt in these two software tools, and describe how they could practically be used. In practice, we rely on the latest public version of \htt\ (\ie\ its version 2 available from the webpage \url{https://www.thphys.uni-heidelberg.de/~plehn/index.php?show=heptoptagger}). Moreover, we have validated our implementations by confronting the results of a few test calculations obtained by using the two interfaced versions of \htt to those returned by \htt when used in a standalone mode.

\subsection{Jet substructure tools in \rivet}\label{sec:rivet}

The implementation of \htt within \rivet has been designed on top of its existing jet-analysis toolkit, using the `smearing projection' machinery that simulates kinematic and particle-identification misreconstruction through transfer functions, while preserving links between particle-level and reconstruction-level physics objects. When jet substructure methods are involved, dedicated smearing methods are required, as many observables (\eg~$N$-subjettinesses) are sensitive to angular correlations between the jet constituents. It is therefore necessary to model the detector's finite angular resolution to get a realistic detector response, including the inefficiencies related to the hadronic calorimeter. This is achieved, as detailed in Ref.~\cite{Buckley:2019stt}, through the directional smearing of the pseudo-rapidity $\eta$ and azimuthal angle $\varphi$ variables defining the direction of every jet constituent. As angular deflections are more significant for constituents with a low transverse momentum $\pt$, this smearing is made $\pt$-dependent with greater angular stability at higher momentum. The specific form used, known to describe jet-substructure effects well on the public data, is angular smearing by a Gaussian with a mean of zero and a standard deviation given by
\begin{eqnarray}
  \sigma_\mathrm{ang} = \frac{\alpha}{1+e^{\beta(p^i_T - \gamma)}} \, .
\end{eqnarray}
Here $\alpha,\ \beta$ and $\gamma$ are free parameters, set to $ 0.045,\ 0.013 $ and $ 31.15 $, respectively, from the fit detailed in Ref.~\cite{Buckley:2019stt}. Additionally, energy-resolution smearing was performed, using relative scaling by a Gaussian with mean of 1, and width $\sigma_E \sim 10\%$.

\begin{table*}
    \centering
    \begin{tabular}{lp{0.65\textwidth}}
      \toprule
      Accessor & Functionality  \\
      \midrule
      \lstinline|const Jet topJet()| & top-quark candidate (returned as a \lstinline|Jet| object)\\
      \lstinline|const Jet bJet()| & $b$-jet candidate (returned as a \lstinline|Jet| object)\\
      \lstinline|const Jet wJet()| & combined subjets compatible with a $W$-boson candidate (returned as a \lstinline|Jet| object)\\
      \lstinline|const Jet w1Jet()| & leading subjet constituting a $W$-boson candidate (returned as a \lstinline|Jet| object)\\
      \lstinline|const Jet w2Jet()| & sub-leading subjet constituting a $W$-boson candidate (returned as a \lstinline|Jet| object)\\
      \lstinline|const PseudoJet topJet()| & top-quark candidate (returned as a \lstinline|PseudoJet| object) \\
      \lstinline|const PseudoJet bJet()| & $b$-jet candidate (returned as a \lstinline|PseudoJet| object)\\
      \lstinline|const PseudoJet wJet()| & combined subjets compatible with a $W$-boson candidate (returned as a \lstinline|PseudoJet| object) \\
      \lstinline|const PseudoJet w1Jet()| & leading subjet constituting a $W$-boson candidate (returned as a \lstinline|PseudoJet| object) \\
      \lstinline|const PseudoJet w2Jet()| & sub-leading subjet constituting a $W$-boson candidate (returned as a \lstinline|PseudoJet| object)\\
      \lstinline|Jets subjets()| & \pt-ordered subjets \\
      \lstinline|double prunedMass()| & pruned top-quark mass \\
      \lstinline|double unfilteredMass()| & mass of the triplet of subjets after unclustering, and before filtering \\
      \lstinline|double deltaTopMass()| & difference between the reconstructed top mass and the true top mass $|m_{\rm rec} - m_t|$ \\
      \lstinline|bool isTopTagged()| & Boolean indicating if the top jet has a mass compatible with the top mass, satisfies two-dimensional mass plane requirements, and has a \pt above some threshold\\
      \lstinline|bool passedMassCutTop()| &  Boolean indicating if the top jet has a mass compatible with the top mass\\
      \lstinline|bool passedMassCut2D()| &  Boolean indicating if the top jet satisfies two-dimensional mass plane requirements\\
      \bottomrule
    \end{tabular}
    \caption{Accessors equipping the \htt wrapper implemented in \rivet.}
    \label{tab:htt_accessors_rivet}
\end{table*}

Our implementation of the \htt method in \rivet relies on an object of the \lstinline{HTT} class, normally to be declared as a member variable of an analysis (or projection) class,
\begin{lstlisting}
 HTT _tagger;
\end{lstlisting}
The \lstinline{HTT} class is defined in the header file \lstinline{"Rivet/Tools/RivetHTT.hh"}. All available parameters for this wrapper are initialised through an \lstinline{HTT::InputParameters} object, that can be used for any further modification relevant for the needs of the user. A simple example is
\begin{lstlisting}
 HTT::InputParameters parameters;
 parameters.mass_drop = 0.8; // mass drop rate
 parameters.filt_N = 5; // nr of filtered subjets
 _tagger.setParams(parameters);
\end{lstlisting}
The list of available parameters can be found in the definition of the C++ structure \lstinline{HTT::InputParameters} in the file \lstinline{"Rivet/Tools/RivetHTT.hh"}. All parameters that are not explicitly initialised by the user keep their default values which have been chosen according to Refs.~\cite{Plehn:2010st, Kasieczka:2015jma}. During the execution of a \rivet analysis, a reclustered jet, instantiated as a \lstinline{Jet} object, can be processed by the methods of the \lstinline{_tagger} object, for example in
\begin{lstlisting}
 _tagger.calc(fatjets[0]);
\end{lstlisting}
Here, \lstinline{fatjets[0]} refers to the leading (\ie~highest-$\pt$) jet included in a vector of clustered jets called \lstinline{fatjets} (the object \lstinline{fatjets} being thus of type \lstinline{Jets}). The computation yields the creation of various accessors that returns a variety of information into native \rivet objects. This list of accessors is shown in Table~\ref{tab:htt_accessors_rivet}.

For a practical example, we refer to the illustrative analysis that can be found in \rivet's \lstinline{"analyses/examples/EXAMPLE_HTT.cc"} file.

\subsection{Jet substructure tools in \ma}\label{sec:ma5}

Since 2021, \ma and its SFS framework for fast simulation of detector effects~\cite{Araz:2020lnp} have been equipped with jet substructure tools and methods.\footnote{See the {\sc GitHub} branch \url{https://github.com/MadAnalysis/madanalysis5/tree/substructure}, as well as the beta versions of \ma v2.0.Z (available at \url{https://github.com/MadAnalysis/madanalysis5/releases}).} In particular, the smearing functionality implemented in the SFS framework allows for modifications of the properties of the jets' constituents, so that the SFS is suitable for the embedding of \htt in a way similar to what was achieved for \rivet in Section~\ref{sec:rivet}. As the substructure branch is so far largely undocumented, we take benefit from the present work to provide some details on its functioning and how to make use of the code to embed top tagging in a generic analysis.

When a jet reconstruction algorithm is turned on in \ma, a so-called `{\it primary}' jet collection is built from a hadronised event. This primary jet collection is equivalent to the sole jet collection that used to be built in versions 1.X.Y of the code, which was documented in~\cite{Conte:2018vmg,Araz:2020lnp}. In practice, the code makes use of its interface with \fastjet~\cite{Cacciari:2011ma}, that can be turned on from the \ma command line interface by typing
\begin{lstlisting}
 set main.fastsim.package = fastjet
\end{lstlisting}
A specific jet algorithm is then activated through the commands
\begin{lstlisting}
 set main.fastsim.algorithm = <algorithm>
 set main.fastsim.<property> = <value>
\end{lstlisting}
The list of supported algorithms, together with the available properties, is provided in~\cite{Araz:2020lnp}. By default, the anti-$k_T$ jet algorithm~\cite{Cacciari:2008gp} is considered, with a radius parameter $R=0.4$ (\lstinline{radius}) and a minimum \pt value of \SI{5}{\GeV} (\lstinline{ptmin}). The primary jet collection is identified by its jet identifier (or \lstinline{JetID}), that is fixed to \lstinline{Ma5Jet} by default. This identifier can be further modified through the command
\begin{lstlisting}
 set main.fastsim.JetID = <new JetID>
\end{lstlisting}

Additional jet collections can be instantiated through
\begin{lstlisting}
 define jet_algorithm <JetID> <algo> [<parameters>]
\end{lstlisting}
where \lstinline{<JetID>} refers to the identifier of the collection, \lstinline{<algo>} to the associated clustering algorithm, and where any algorithm-specific parameter can be optionally fixed through comma-separated or space-separated equalities (otherwise default values are used). For instance, typing
\begin{lstlisting}
 define jet_algorithm CA08 cambridge radius=0.8 \
    ptmin=200
\end{lstlisting}
defines a jet collection coined \lstinline{CA08}, in which jets are reconstructed by means of the Cambridge-Aachen jet algorithm~\cite{Dokshitzer:1997in, Bentvelsen:1998ug, Wobisch:1998wt}, with a radius parameter set to 0.8 and a minimum \pt value of \SI{200}{\GeV}. Parameters can also be altered through specific commands, like for instance in
\begin{lstlisting}
 set CA08.radius = 0.8
 set CA08.ptmin = 200
\end{lstlisting}
Once multiple jet collections are defined, constituent-based smearing is always applied to the properties of all final-state hadrons before the different reconstructions are performed. This contrasts with the setup in which a single collection is defined, as here users can decide to smear reconstructed objects instead of their constituents. Reconstruction efficiencies can also be provided from the command line interface (see~\cite{Araz:2020lnp}), but they will only be applied to the primary jet collection. This limiting behaviour can however be bypassed by employing the expert mode of the code, in which users implement their analysis directly in C++ (and are thus free to do whatever they want). We therefore focus only on this expert mode from now on.\footnote{The command line interface of \ma can be used for the generation of a skeleton C++ code, that can then be modified in a second step according to the wishes of the user.}
\begin{table*}
    \centering
    \begin{tabular}{lp{0.65\textwidth}}
      \toprule
      Accessor & Functionality  \\
      \midrule
      \lstinline|const RecJet top()| & top-quark candidate (returned as a \lstinline|RecJet| object)\\
      \lstinline|const RecJet b()| & $b$-jet candidate (returned as a \lstinline|RecJet| object)\\
      \lstinline|const RecJet W()| & combined subjets compatible with a $W$-boson candidate (returned as a \lstinline|RecJet| object)\\
      \lstinline|const RecJet W1()| & leading subjet constituting a $W$-boson candidate (returned as a \lstinline|RecJet| object)\\
      \lstinline|const RecJet W2()| & sub-leading subjet constituting a $W$-boson candidate (returned as a \lstinline|RecJet|object)\\
      \lstinline|RecJets subjets()| & \pt-ordered subjets (returned as a vector of \lstinline|RecJet| objects, also known as a \lstinline|RecJets| object)\\
      \lstinline|MAfloat32 pruned_mass()| & pruned top-quark mass \\
      \lstinline|MAfloat32 unfiltered_mass()| & mass of the triplet of subjets after unclustering, and before filtering \\
      \lstinline|MAfloat32 delta_top()| & difference between the reconstructed top mass and the true top mass $|m_{\rm rec} - m_t|$ \\
      \lstinline|MAbool is_tagged()| & Boolean indicating if the top jet has a mass compatible with the top mass, satisfies two-dimensional mass plane requirements, and has a \pt above some threshold\\
      \lstinline|MAbool is_maybe_top()| & Boolean indicating if the top jet has a mass compatible with the top mass\\
      \lstinline|MAbool is_masscut_passed()| & Boolean indicating if the top jet satisfies two-dimensional mass plane requirements\\
      \bottomrule
    \end{tabular}
    \caption{Accessors equipping the HEPTopTagger wrapper implemented in \ma.}
    \label{tab:htt_accessors}
\end{table*}

At the level of the C++ code generated by \ma (or implemented from scratch by expert users), the primary jet collection can be accessed through the standard accessor \lstinline{event.rec()->jets()} (as described in~\cite{Conte:2014zja, Conte:2018vmg}), and all jet collections (including the primary one) can be accessed through the accessor \lstinline{event.rec()->jets(<JetID>)} (with \lstinline{<JetID>} being the identifier referring to the collection). These accessors return a vector of pointers to constant \lstinline{RecJetFormat} objects (or \lstinline{RecJet} objects for short), the entire vector being also of the shorthand type \lstinline{RecJets}.

In the version 2.0.X of \ma, a \lstinline{Substructure} namespace has been implemented and includes wrappers to a large set of \fastjet and \fjcontrib functionalities. This substructure module allows for three standard infrared and collinear safe jet-clustering algorithms, that can be initialised as for instance through
\begin{lstlisting}
 Substructure::Cluster cluster;
 cluster.Initialize(Substructure::antikt, 0.4, 20., isExclusive = false);
\end{lstlisting}
This initialises a \lstinline{Cluster} object named \lstinline{cluster} in which jet reconstruction relies on the anti-$k_T$ algorithm with parameter $R=0.4$, and that selects reconstructed jets featuring $\pt > \SI{20}{\GeV}$. In order to make use of the Cambridge-Aachen or the generalised $k_T$~\cite{Cacciari:2011ma} algorithm, the first argument of the \lstinline{Initialize} method needs to be set to \lstinline{Substructure::cambridge} and \lstinline{Substructure::kt} respectively. The next arguments are related to the two options available for the three supported algorithms (namely the radius parameter $R$ and the minimum \pt requirement applied on the reconstructed jets), and the last optional argument (\lstinline{isExclusive}) indicates whether leptons and photons originating from hadron decays have to be included in their respective collections in addition to be considered as jet constituents (\lstinline{isExclusive = false}), or not (\lstinline{isExclusive = true}). Next, clustering is executed through the command
\begin{lstlisting}
 cluster.Execute(<event>, <JetID>);
\end{lstlisting}
where \lstinline{<JetID>} is the identifier of the jet collection to use to store the output of the clustering, and \lstinline{<event>} is an \lstinline{EventFormat} object pointing to the whole event. Smearing and reconstruction efficiencies are automatically included, if provided by the user (see Ref.~\cite{Araz:2020lnp}).

Clustered jets can be further manipulated, either one by one or all together. For instance, the first of the following commands defines a new collection \lstinline{FilteredJets} as a sub-selection of all reconstructed (primary) jets satisfying $\pt > \SI{20}{\GeV}$ and $|\eta| < 2.5$. The next two lines are dedicated to the initialisation of a new clustering method (the Cambridge-Aachen algorithm with a radius parameter $R=0.5$, that is the sole parameter that can be specified here), with which those jets will be reclustered,
\begin{lstlisting}
 RecJets FilteredJets = filter(event.rec()->jets(), 20.0, 2.5);
 Substructure::Recluster recluster;
 recluster.Initialize(Substructure::cambridge, 0.5);
\end{lstlisting}
Here, we assume that the primary jets have been clustered through some (unspecified) algorithm. Next, we make use of the \lstinline{Recluster} object, a first time on the whole jet collection, and a second time specifically on the leading jet,
\begin{lstlisting}
 RecJets ReclusteredJets = recluster.Execute(FilteredJets);
 const RecJet ReclusteredLeadingJet = recluster.Execute(FilteredJets[0]);
\end{lstlisting}

As another example, we now discuss jet reconstruction in which the radius parameter $R$ is variable~\cite{Krohn:2009zg}.\footnote{See the webpage \url{https://phab.hepforge.org/source/fastjetsvn/browse/contrib/contribs/VariableR/tags/1.2.1/} for more information.} Such a method can be used from the \lstinline{Substructure} wrapper as follows,
\begin{lstlisting}
 Substructure::VariableR variableR;

 MAfloat32 rho=2000., minR=0., maxR=2., ptmin=0.0;
 Substructure::VariableR::ClusterType clusterType =
   Substructure::VariableR::AKTLIKE;
 Substructure::VariableR::Strategy strategy =
   Substructure::VariableR::Best;
 MAbool isExclusive = false;

 variableR.Initialize(rho, minR, maxR, clusterType, strategy, ptmin, isExclusive);
\end{lstlisting}
The clustering type must be \lstinline{CALIKE} (Cambridge-Aachen), \lstinline{KTLIKE} ($k_T$ algorithm) or \lstinline{AKTLIKE} (anti-$k_T$ algorithm), the parameters \lstinline{minR} and \lstinline{maxR} stand for the minimum and maximum radius values allowed, and the internal clustering strategy to be used by \fastjet has to be among \lstinline{Best}, \lstinline{N2Tiled}, \lstinline{N2Plain}, \lstinline{NNH} or \lstinline{Native}. We refer to Ref.~\cite{Krohn:2009zg} for more information. Reclustering is then proceeded as above,
\begin{lstlisting}
 RecJets variableRJets = variableR.Execute(FilteredJets);
 RecJets variableRLeadingJet = variableR.Execute(FilteredJets[0]);
\end{lstlisting}

In order to enable the usage of \htt within \ma, the package must first be downloaded and linked to the code. This is achieved by typing in the \ma command line interface
\begin{lstlisting}
 install HEPTopTagger
\end{lstlisting}
once \fastjet and \fjcontrib are installed and available (which is achieved by typing in the interpreter the command \lstinline{install fastjet}). When implementing an analysis in C++, the execution of \htt is controlled from a dedicated structure called \lstinline{Substructure::HTT::InputParameters}. The latter is defined in the file \lstinline{"tools/SampleAnalyzer/Interfaces/HEPTopTagger/HTT.h"}, together with all associated parameters and methods, and it is documented in the file \lstinline{"tools/SampleAnalyzer/Interfaces/HEPTopTagger/README.md"}. Taking the example introduced in Section~\ref{sec:rivet}, a simple example of initialisation would read
\begin{lstlisting}
 Substructure::HTT::InputParameters parameters;
 parameters.mass_drop = 0.8;
 parameters.filt_N = 5;

 Substructure::HTT tagger;
 tagger.Initialize(parameters);
\end{lstlisting}
\htt is then executed as
\begin{lstlisting}
 tagger.Execute(filteredJets[0]);
\end{lstlisting}
As for the embedding into \rivet, this method leads to the generation of a variety of accessors that allows for the exploration of the properties of the would be top-jet.  Their list is given in Table~\ref{tab:htt_accessors}. For more detailed practical examples on the usage of jet substructure techniques and \htt within \ma, we refer to the tutorial available from \url{https://github.com/MadAnalysis/tutorial_osu}.

\section{Exploring new physics effects with boosted top quarks in the SMEFT} \label{sec:top_eft}
In this section, we demonstrate the use of \htt (version 2) within the \rivet and \ma frameworks, and we study the potential impact of SMEFT operators on boosted top quark decays. The set of relevant operators that we consider is introduced in Section~\ref{sec:topeft_theo}. In Section~\ref{sec:SMEFT_perf}, we focus on the production of a semi-leptonically decaying $\ttbar$ pair to investigate how SMEFT deviations in the properties of boosted top quarks affect the performance of top taggers (through deviations from the taggers' expectations of SM-like top-quark decay properties). Next, we make use of our findings to derive in Section~\ref{sec:collider} the sensitivity of a typical analysis of boosted top-pair production and decay to various SMEFT operators poorly constrained by other means.

\subsection{Theoretical framework}\label{sec:topeft_theo}
In the absence of any explicit evidence for new fields and interactions beyond the SM, effective field theories provide a natural path to scrutinising the impact of hypothetical BSM physics at the electroweak scale $\Lambda_{\rm EW}$. In this context, the SMEFT paradigm offers a very promising framework allowing for the exploration of heavy new physics. The SMEFT is an effective field theory expansion in an energy scale $\Lambda$ that is assumed to satisfy $\Lambda \gg \Lambda_{\rm EW}$. The model Lagrangian is defined via a set $\{ {\cal O}_1, {\cal O}_2, ... \}$ of higher-dimensional (\ie~non-renormalisable) operators in the SM fields. Assuming that the leading new-physics effects arise at dimension six, this Lagrangian reads
\begin{equation}\label{eq:lagsmeft}
  \mathcal{L}_\mathrm{SMEFT} = {\cal L}_{\rm SM} + \sum_j \frac{C_j}{\Lambda^2}\mathcal{O}_j + {\cal O}\big(1/\Lambda^3\big)\,,
\end{equation}
where $\mathcal{L}_\mathrm{SM}$ is the SM Lagrangian, and the Wilson coefficients $C_j$ encode the BSM details of the theory. Among the \num{3045} free parameters in this general SMEFT Lagrangian of eq.~\eqref{eq:lagsmeft}~\cite{Grzadkowski:2010es, Henning:2015alf}, only a few are relevant for top-quark physics.

We consider a scenario in which $CP$ is conserved, and we next assume that new physics only couples to the weak doublet of left-handed top and bottom quarks ($Q$) and the right-handed weak singlets ($t$ and $b$) of third-generation quarks (as well as to SM bosons). Moreover, bosonic operators leading to flavour-universal effects are discarded, we approximate the CKM matrix by the identity matrix, and all Yukawa couplings but those of the top and bottom quarks are neglected. In order to further reduce the number of free parameters, we consider a $U(2)_q\times U(2)_u \times U(2)_d$ flavour symmetry among the quarks of the first and second generations, in agreement with the principle of minimal flavour violation~\cite{Chivukula:1987py, Hall:1990ac, DAmbrosio:2002vsn}. Differences between the first and second-generation quarks are thus ignored, and we subsequently introduce the generic notation $q$ for a left-handed weak doublet of first-generation or second-generation quark fields, and $u$ and $d$ for the corresponding right-handed weak singlets of up-type and down-type quark fields.

In our analysis, we aim to leverage the detector-simulation capabilities of the \ma and \rivet frameworks (including our implementation of \htt) to realistically explore the effects of effective operators on the reconstruction performance of boosted top quarks. Among the full set of potentially impactful SMEFT operators~\cite{AguilarSaavedra:2018nen}, only eight of them are not too strongly constrained by other means~\cite{Buckley:2015lku, Hartland:2019bjb, Brivio:2019ius, Ellis:2020unq, Ethier:2021bye, Giani:2023gfq}, so that an investigation of pair-production and decay of boosted top quarks could offer new handles on them. They read, in the notation of Ref.~\cite{Brivio:2019ius},
\begin{equation}\begin{split}
  \mathcal{O}_{Qq}^{1,8} =& (\bar{Q} \gamma_\mu T^A Q)\ \ (\bar{q} \gamma^\mu T^A q)\,,
  \\
  \mathcal{O}_{Qq}^{3,8} =& (\bar{Q} \gamma_\mu T^A\sigma^I Q)\ \ (\bar{q} \gamma^\mu T^A \sigma^I q)\,,\\
  \mathcal{O}_{Qq}^{3,1} =& (\bar{Q} \gamma_\mu\sigma^I Q)\ \ (\bar{q} \gamma^\mu\sigma^I q)\,,
  \\
  \mathcal{O}_{tu}^8 =& (\bar{t}\gamma_\mu T^A t)\ \ (\bar{u}\gamma^\mu T^A u)\,,\\
  \mathcal{O}_{td}^8 =& (\bar{t}\gamma_\mu T^A t)\ \ (\bar{d}\gamma^\mu T^A d)\,,
  \\
  \mathcal{O}_{Qu}^8 =& (\bar{Q}\gamma_\mu T^A Q)\ \ (\bar{u}\gamma^\mu T^A u) \,,\\
  \mathcal{O}_{Qd}^8 =& (\bar{Q}\gamma_\mu T^A Q)\ \ (\bar{d}\gamma^\mu T^A d)\,,
  \\
  \mathcal{O}_{tq}^8 =& (\bar{q}\gamma_\mu T^A q)\ \ (\bar{t}\gamma^\mu T^A t)\,,
\end{split}\label{eq:ops}\end{equation}
where the matrices $T^A$ stand for the generators of $SU(3)_c$ in the fundamental representation, and the matrices $\sigma^I$ are the usual Pauli matrices.

\subsection{Top tagging performance in the presence of non-vanishing SMEFT operators}\label{sec:SMEFT_perf}

In order to assess how non-zero values for the Wilson coefficients associated with the SMEFT operators of eq.~\eqref{eq:ops} affect top-quark tagging performance, we make use of \madgraph version~3.0.3~\cite{Alwall:2014hca} to generate parton-level events describing top-antitop production and their semi-leptonic decay at the LHC (operating at a centre-of-mass energy of \SI{13}{\TeV}). We rely on leading-order matrix elements convolved with the leading-order set of NNPDF3.0 parton distribution functions~\cite{Ball:2014uwa} provided through the \textsc{Lhapdf6} library~\cite{Buckley:2014ana}. For efficiency reasons, the Monte Carlo event generation was kinematically biased to high scales, and we required that the invariant mass of the produced \ttbar system satisfies $m_{t\bar t}^{\rm truth} > \SI{950}{\GeV}$. These fixed-order events are matched with parton showering and hadronisation as modelled by \pythia\ version 8.2~\cite{Sjostrand:2014zea}. Background events are generated with the same toolchain, but by considering the production of a leptonically-decaying $W$ boson in association with a pair of $b$ jets (and two additional jets), $pp\to W b \bar b + \text{jets}$.

Our canonical analysis was implemented in \rivet version~3~\cite{Bierlich:2019rhm}.\footnote{An equivalent implementation in \ma produced similar results.} It employs \fastjet version 3.3.3~\cite{Cacciari:2011ma} for event reconstruction, and \htt version 2~\cite{Kasieczka:2015jma} in its default configuration. We remind that the latter has been tuned on boosted top quarks with properties as expected from their SM production and decay, which may thus not be the best for scenarios in which SMEFT effects change the properties of the produced tops. In our usage of \htt, we turn on the `optimal $R$' option. This allows the tagging algorithm to determine the minimum choice for the fat jet reconstruction radius to ensure that the reconstructed top jet includes a three-prong structure (as expected from standard top-quark decays).

Our event reconstruction is achieved by first defining a collection of `small jets' through the clustering of all visible hadron-level final-state objects with a pseudo-rapidity $|\eta|<4.5$, muons excepted. We use the \akt jet algorithm~\cite{Cacciari:2008gp} with radius parameter $R=0.4$, and then impose a minimum transverse-momentum requirement of $\pt > \SI{30}{\GeV}$ on the reconstructed small jets. Next, we define a collection of `fat jets' from the same hadron-level objects. This collection is constructed by using the Cambridge-Aachen algorithm~\cite{Dokshitzer:1997in, Bentvelsen:1998ug, Wobisch:1998wt} with a radius parameter $R=1.5$. We impose a minimum transverse momentum requirement of $\pt > \SI{200}{\GeV}$ on the reconstructed fat jets.

Lepton candidates (\ie~electrons and muons) are required to satisfy basic momentum and pseudo-rapidity criteria, $\pt > \SI{10}{\GeV}$ and $|\eta|<2.5$. At this stage, $\Delta R$-based isolation is enforced in order to remove the overlap between the lepton collection and the two jet collections. We remove from the small-jet collection any small jet $j$ lying in the vicinity of a lepton $\ell$ by an angular distance $\Delta R(\ell,j) < 0.1$, and we then discard any lepton lying at a distance $\Delta R(\ell,j) < 0.4$ of any of the remaining small jets. Moreover, we define $b$ jets as small jets with $\pt > \SI{30}{\GeV}$ and with a ghost-associated $b$-hadron with $\pt > \SI{5}{\GeV}$~\cite{Cacciari:2007fd,Cacciari:2008gn}.

After reconstruction, we select events whose topology is compatible with that expected from the production of a pair of boosted top quarks that decays semi-leptonically. We require that each selected event features one lepton with at least \SI{50}{\GeV} of transverse momentum, a minimum missing transverse energy $\slashed{E}_T > \SI{30}{\GeV}$, as well as at least two small $b$ jets and two small light jets. Next, we reconstruct the leptonically-decaying $W$ boson that we consider on-shell. This assumption implies that the invariant mass of the system comprising the lepton and the missing momentum is equal to the mass $m_W$ of the $W$ boson, which allows us to determine the longitudinal component $\slashed{p}_z$ of the missing momentum,
\begin{equation}\label{eq:wreco}\begin{split}
  m^2_W =&\
  (\mathbf{p}_\ell + \slashed{\mathbf{p}})^2 \\ =&\
  \Big(E_\ell+\sqrt{\slashed{p}^2_T+\slashed{p}^2_z}\Big)^2 -
  \sum_i\Big(p_{\ell, i} + \slashed{p}_i\Big)^2 \\
 & \quad \text{with}\ i \in \{x,\ y,\ z\}\,.
\end{split}\end{equation}
In the above expression, $\mathbf{p}_\ell = (p_{\ell, x}, p_{\ell, y}, p_{\ell, z})$ denotes the three-momentum of the lepton, $\slashed{\mathbf{p}} = (\slashed{p}_x, \slashed{p}_y, \slashed{p}_z)$ is the missing three-momentum, and $E_\ell$ stands for the energy of the lepton. From the solution to eq.~\eqref{eq:wreco}, we can define the four-momentum of the leptonically-decaying $W$ boson \WL. In the case where this equation has two solutions, we arbitrarily choose the smallest value for $\slashed{p}_z$. Moreover, when it has no solution, we set the associated discriminant to 0 and use the resulting solution.

In order to reconstruct the leptonically-decaying top quark, we match this reconstructed $W$ boson with one of the $b$ jets by minimising the difference between the top mass $m_t$ and the invariant mass $m[\WL\oplus b]$ of the system constituted of the reconstructed $W$ boson \WL and the $b$ jet. This is achieved through a $\Delta\chi^2$ minimisation,
\begin{equation}\label{eq:tLchi2}
  \Delta\chi^2
    =      \frac{|m_t - m[\WL\oplus b]|^2}{\sigma^2}
    \equiv \frac{|m_t - m(\tL)|^2}{\sigma^2}\,,
\end{equation}
with a mass-resolution parameter $\sigma = \SI{40}{\GeV}$. The $b$-jet matched in this leptonic-top reconstruction is denoted by $\bL$ in the following text.

\begin{figure}
  \centering
  \includegraphics[width=0.48\textwidth]{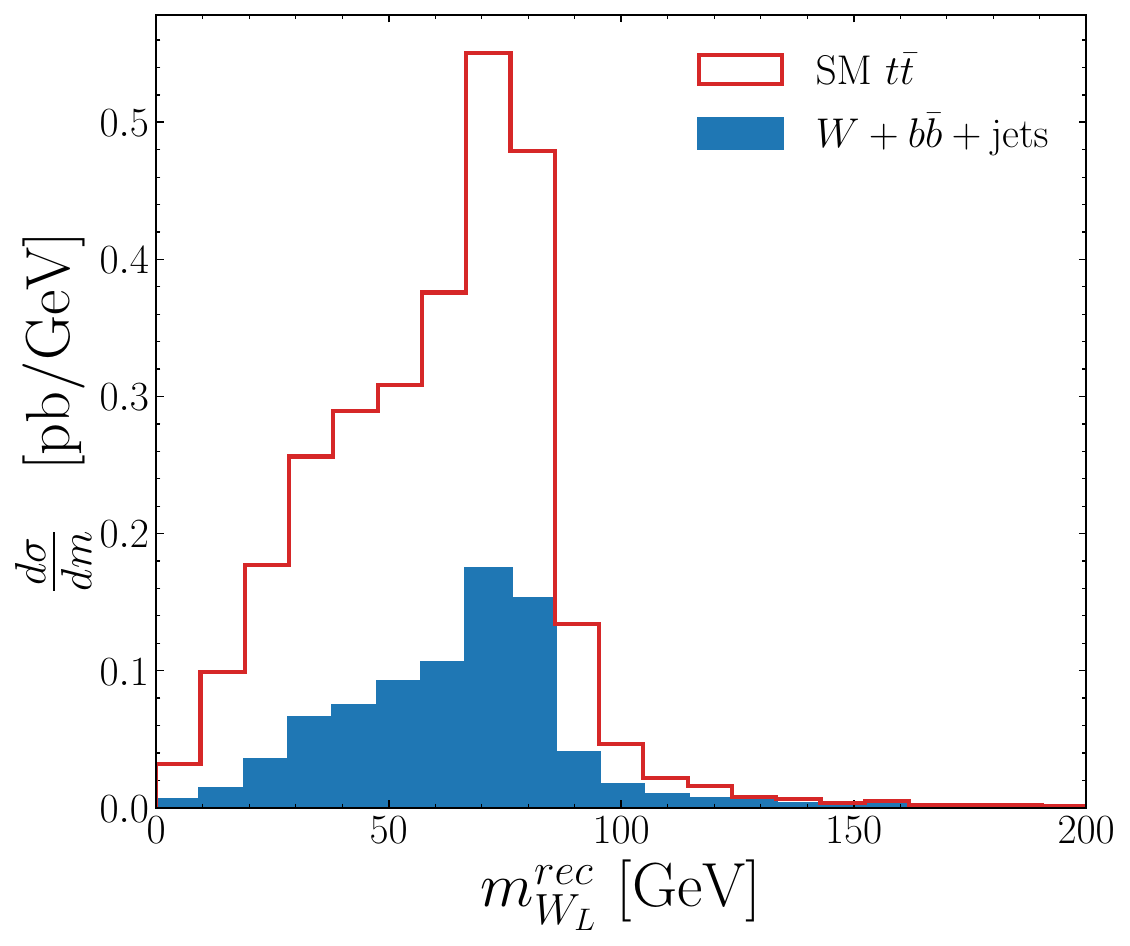}
  \includegraphics[width=0.48\textwidth]{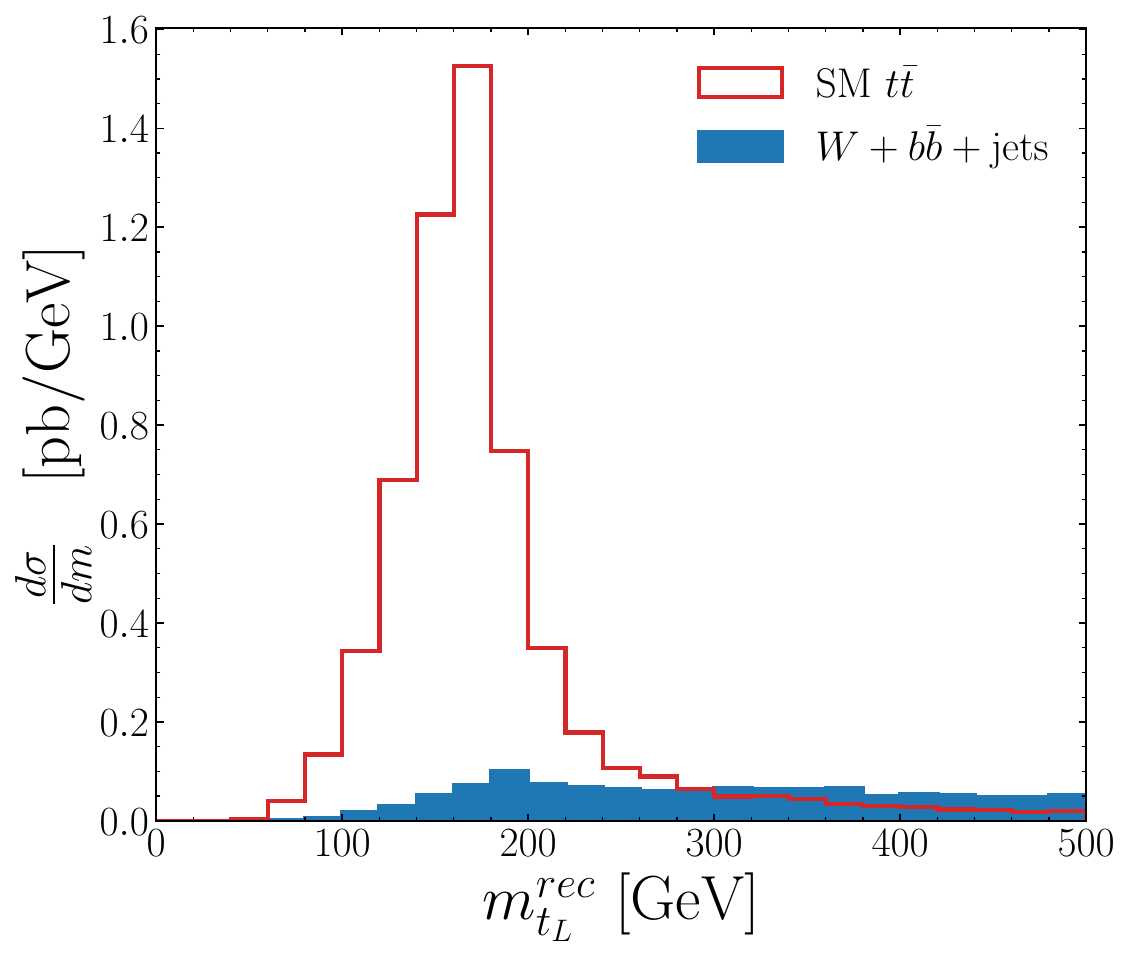}
  \caption{Invariant mass spectra relevant to the reconstruction of the leptonically decaying top quark. We display the invariant mass $m(\WL)$ of the reconstructed $W$ boson (upper panel), as well as that ($m(\tL)$) of the reconstructed top quark (lower panel). Predictions are shown for both the $\ttbar$ signal (red) and the associated background (blue).}
  \label{fig:leptonic_reco}
\end{figure}

Figure~\ref{fig:leptonic_reco} illustrates the features of the reconstruction of the leptonic branch of the process. It shows the distribution in the invariant mass $m(\WL)$ of the reconstructed $W$ boson (upper panel) and that in the invariant mass $m(\tL)$ of the reconstructed top quark (lower panel). Predictions are displayed both for the \ttbar signal (red) and the associated background (blue). These results demonstrate that most signal events exhibit an on-shell leptonically-decaying $W$-boson and an on-shell associated top quark. However, the tails of the distributions extend quite significantly away from the peak values for the two spectra. This originates from the inefficiencies inherent to the kinematic fit performed in eq.~\eqref{eq:wreco}, which could lead to zero, one, or two solutions for $\slashed{p}_z$. Consequently, the reconstructed mass of the \WL boson (upper panel of Figure~\ref{fig:leptonic_reco}) exhibits a plateau at values lower than the true $W$ mass. This impacted our choice for the numerical value of the resolution parameter used in the $\chi^2$ fit of eq.~\eqref{eq:tLchi2}, which then leads to a quite broad peak around the true top mass for the distribution in the reconstructed top mass (lower panel of Figure~\ref{fig:leptonic_reco}).

In the next step of our analysis, we study to which extent a hadronically-decaying top quark can be reconstructed from the event's final state. We start from the fat-jet collection and discard any fat jet $J$ that lies at angular distance $\Delta R (J, \tL) \leq 1.5$ of the reconstructed leptonically-decaying top quark $\tL$. Next, we discard all fat jets found near the $\bL$ jet, \ie~lying within a angular distance $\Delta R (J, \bL) \leq 1.5$. Finally, we reject events that do not comprise at least one fat jet that includes a (small) $b$-jet. This condition is implemented by requiring that there is a fat jet $J$ such that a $b$-jet different from the $\bL$ jet lies at a distance $\Delta R(J,b) < 1.5$ from it. We then test whether the hardest of the remaining fat jet is top-tagged by \htt.

\begin{figure*}
  \centering
  \includegraphics[scale=0.5]{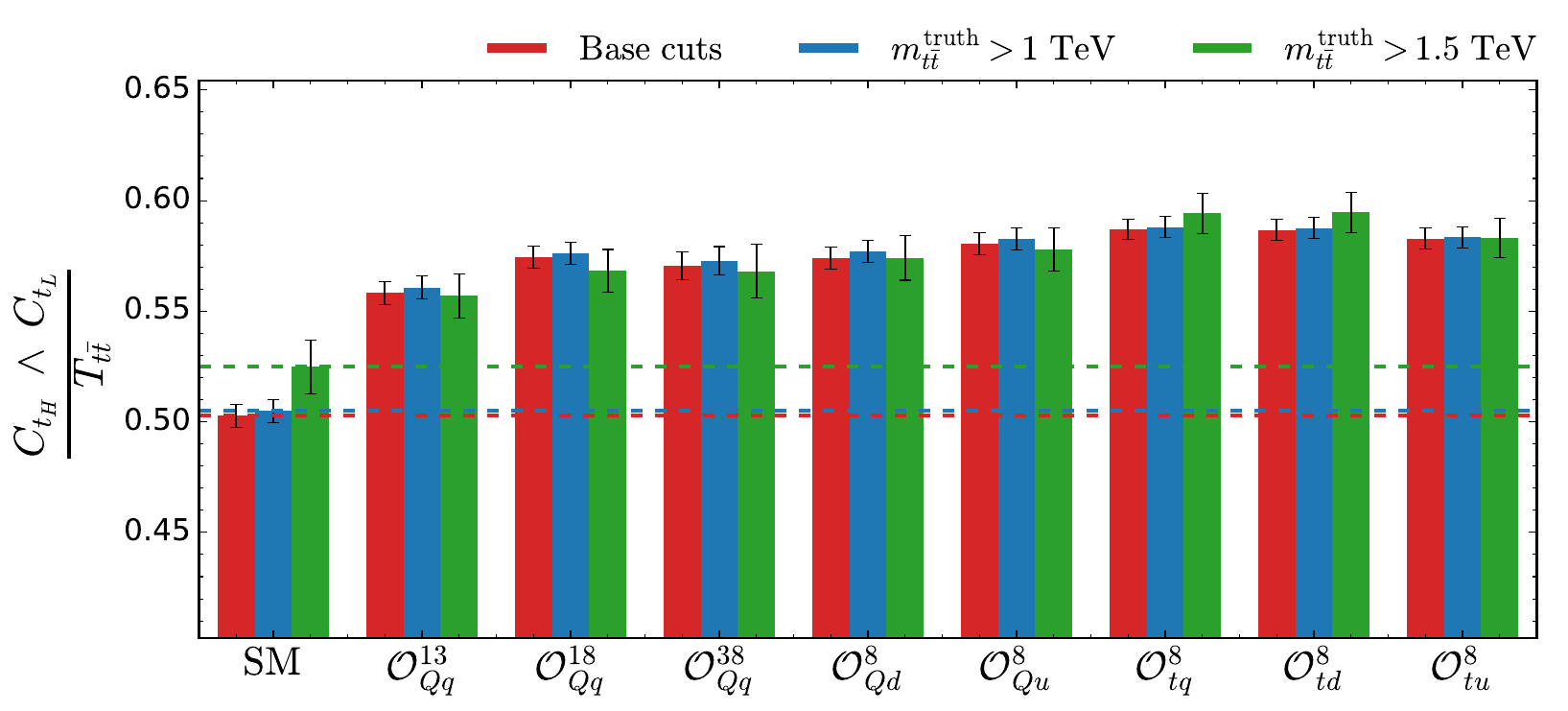}
  \caption{Efficiency associated with the reconstruction of one leptonic and one hadronic top quark, estimated relatively to the number of events containing two on-shell top quarks. Results are shown after the analysis baseline cuts (red), an additional $m^\text{truth}_\ttbar > \SI{1}{\TeV}$ cut (blue), and an extra $m^\text{truth}_\ttbar > \SI{1.5}{\TeV}$ cut (green). We consider the case of the SM (first column), as well as when eight different SMEFT operators are turned on (next columns).}
  \label{fig:quality_notsmeared_1}
\end{figure*}

We now introduce a few useful quantities in order to assess the performance of \htt. First, we classify a truth-level top quark as ``on-shell'' when its invariant mass is in the range $[m_t-\SI{15}{\GeV}, m_t+\SI{15}{\GeV}]$, and define the quantity $T_{\ttbar}$ as the number of \ttbar events featuring two such on-shell top quarks. Next, we denote by $C_{t_H}$ the number of events for which the reconstructed hadronic top quark lies within an angular distance $\Delta R <1.2$ from the corresponding truth-level object when the latter is on-shell.\footnote{In our notation, $T$ is related to `true' and $C$ to `corresponding'.} Similarly, $C_{\tL}$ stands for the number of events for which the reconstructed leptonically-decaying top quark lies at a distance $\Delta R<1.2$ of its truth-level counterpart when it is on-shell. The quantities $C_{t_H}$ and $C_{\tL}$ hence refer to the number of events for which the reconstructed top quarks are matched with the corresponding truth-level objects so that reconstruction is deemed correct.

With the first set of three coloured columns displayed on the left of Figure~\ref{fig:quality_notsmeared_1}, we show the resulting reconstruction efficiency defined as the ratio of the number of events featuring correctly reconstructed hadronic and leptonic top quarks to the number of events including two truth-level on-shell top quarks, \ie~the self-explanatory quantity
\begin{equation}
    \varepsilon = 
      \frac{C_{t_H}~\land~C_{\tL}}{T_{\ttbar}} \equiv 
      \frac{\left(C_{t_H}~\text{\textsc{and}}~C_{\tL}\right)}{T_{\ttbar}}\,.
\end{equation}
This efficiency is given when the baseline cuts described above are imposed (red), when an additional selection of $m^\text{truth}_\ttbar > \SI{1}{\TeV}$ is enforced (blue), and finally, when we require $m_{\ttbar}^\text{truth} > \SI{1.5}{\TeV}$ (green). The error bars represent the related Monte Carlo statistical uncertainty. We observe that about 50\% of the SM events with on-shell \ttbar production are correctly reconstructed, this number slightly increasing when we focus more deeply on the boosted regime (\ie~with a larger $m^\text{truth}_\ttbar$ cut).

\begin{figure*}
  \centering
  \includegraphics[scale=0.5]{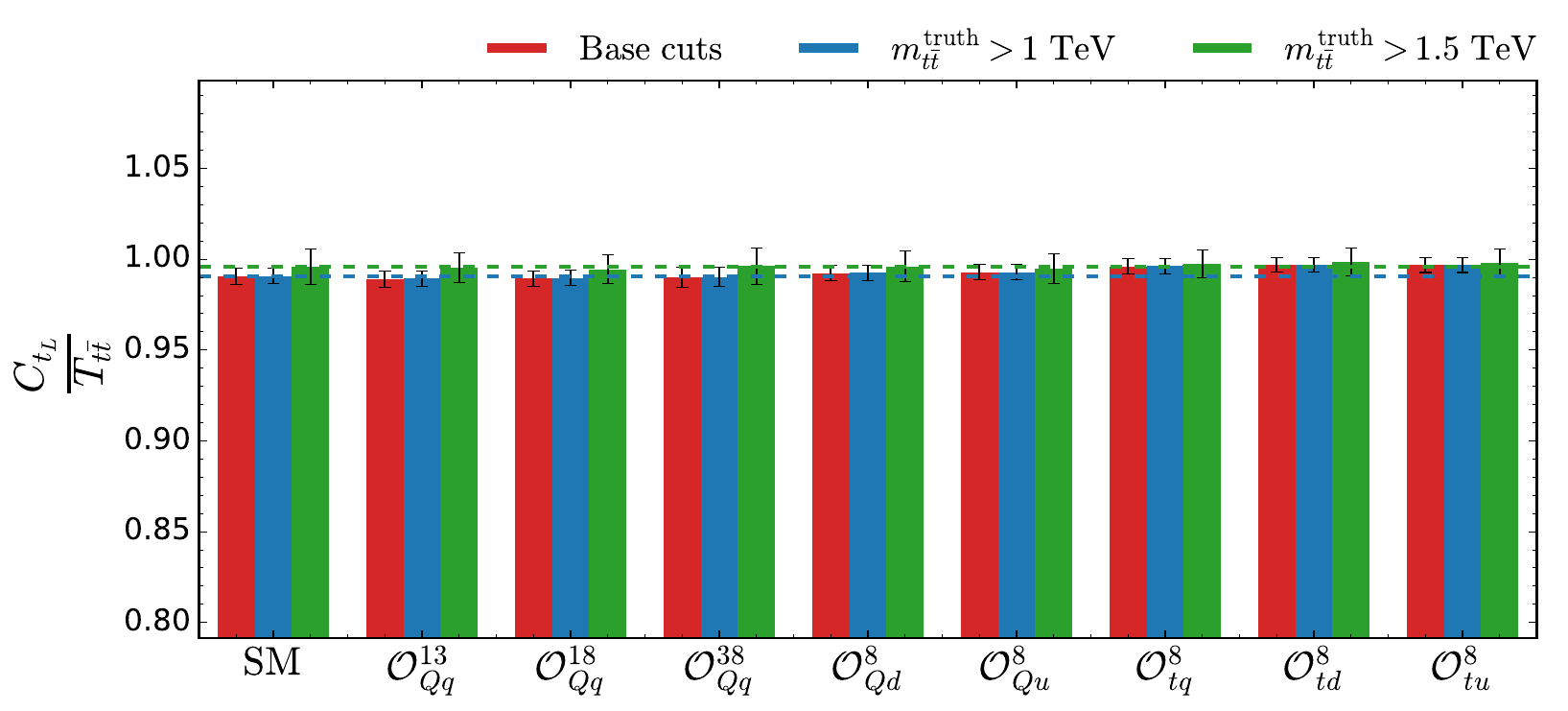}
  \caption{Same as in Figure~\ref{fig:quality_notsmeared_1} but for the efficiency associated with the reconstruction of one leptonic top quark, estimated relatively to the number of events containing two on-shell top quarks.}\label{fig:quality_notsmeared_2}
\end{figure*}

\begin{figure*}
  \centering
  \includegraphics[width=0.48\textwidth]{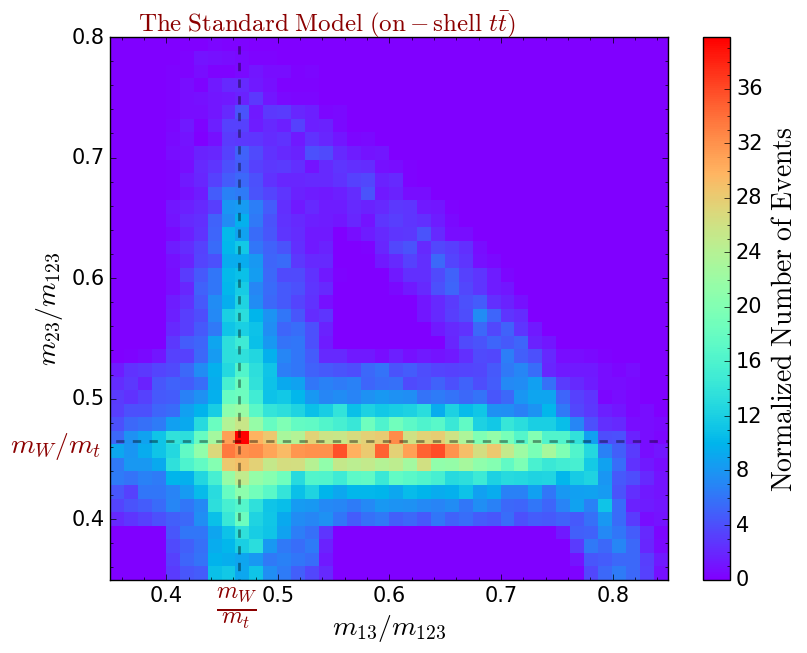}
  \includegraphics[width=0.48\textwidth]{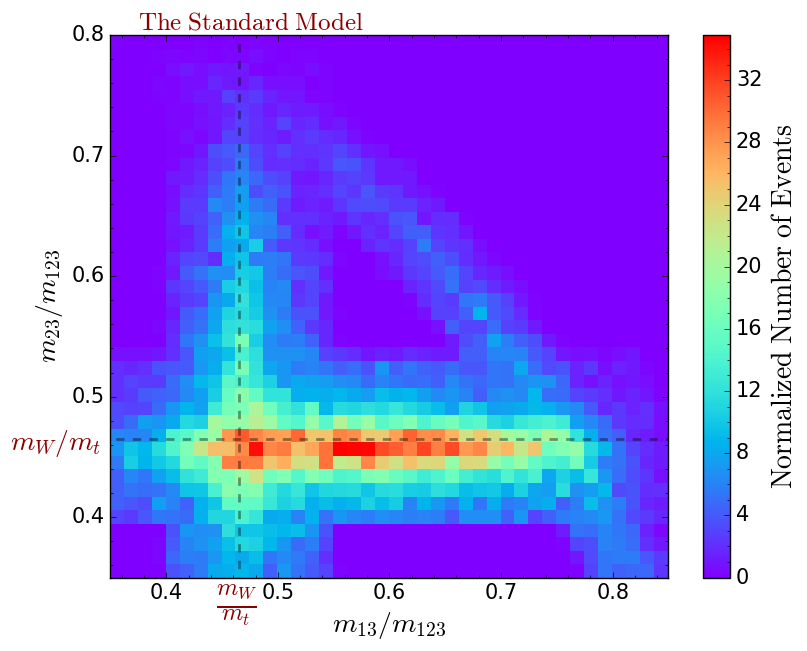}\\
  \includegraphics[width=0.48\textwidth]{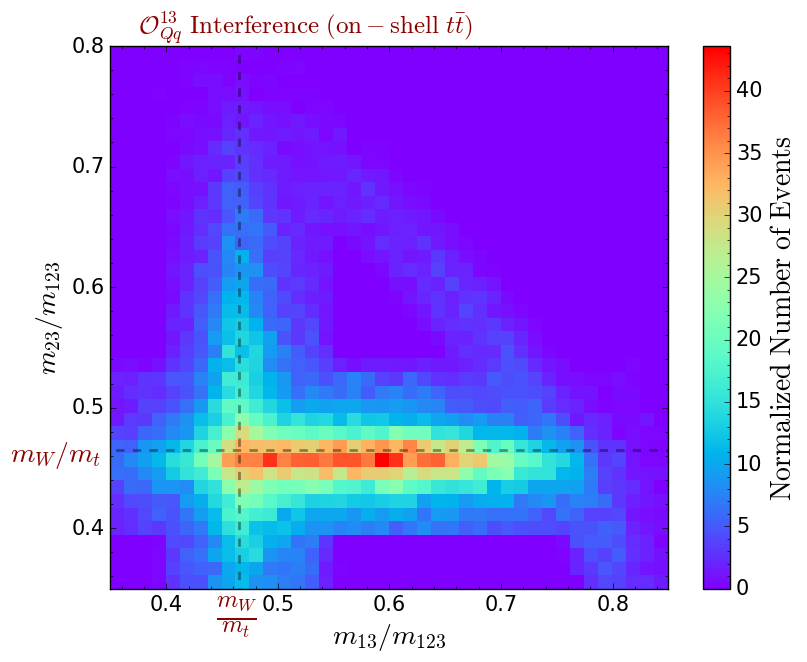}
  \includegraphics[width=0.48\textwidth]{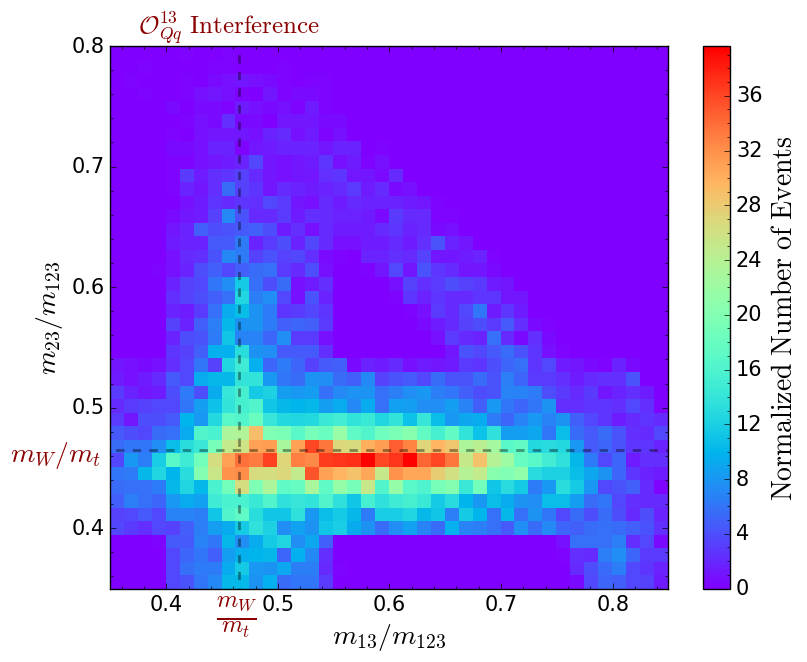}\\
  \includegraphics[width=0.48\textwidth]{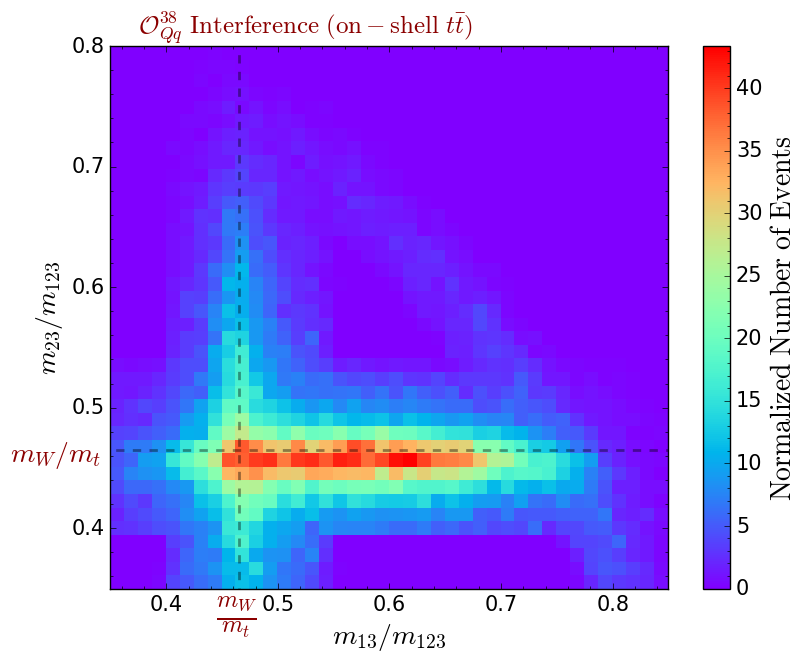}
  \includegraphics[width=0.48\textwidth]{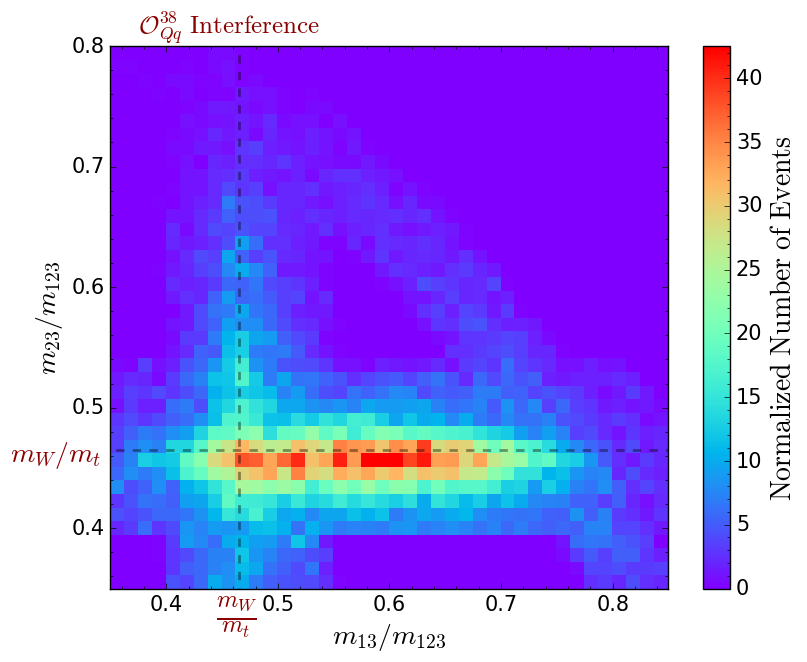}\\
  \caption{Dalitz plots depicting the invariant mass ratios $m_{13}/m_{123}$ and $m_{23}/m_{123}$ where the indices refer to a specific jet among those comprising the reconstructed hadronically-decaying top quark. We show predictions when the on-shellness of the top quark is enforced (left column) and when there is no restriction on the invariant mass of the top quarks at parton level (right column). We consider the case of the SM (top row) and that of scenarios with one SMEFT operator turned on, namely ${\cal O}_{Qq}^{3,1}$ (middle row) and ${\cal O}_{Qq}^{3,8}$ (bottom row).
    \label{fig:dalitz}}
\end{figure*}

The efficiency, however, increases once one of the SMEFT operators of eq.~\eqref{eq:ops} is turned on, as shown in the rest of Figure~\ref{fig:quality_notsmeared_1} (the dashed lines being guidelines for the comparison with the case of the SM). Here, the signal is simulated by implementing the Lagrangian and operators of eqs.~\eqref{eq:lagsmeft} and~\eqref{eq:ops} in \textsc{FeynRules} as specified in Refs.~\cite{Christensen:2009jx, Alloul:2013bka}. This is then used to generate a UFO~\cite{Degrande:2011ua} model to be used within \madgraph so that events could be generated through the same toolchain as that described at the beginning of this section. However, whereas we include the interference of dimension-six contributions with SM diagrams, squared SMEFT contributions (thus formally of dimension-eight) are truncated away. The increase in efficiency observed in Figure~\ref{fig:quality_notsmeared_1} can be traced back not only to a slight increase in the signal cross section, but also to a change in the event topology enhancing \htt's ability to correctly tag the boosted, hadronically-decaying top quark. To prove this statement, we display in Figure~\ref{fig:quality_notsmeared_2} the efficiency $\varepsilon'$ of correctly tagging the leptonic top $\tL$ regardless of the hadronic branch of the events,
\begin{equation}
    \varepsilon' = \frac{C_{\tL}}{T_{\ttbar}}\,.
\end{equation}
As can be seen in this figure, the efficiency  $\varepsilon'$ is almost 100\% for all considered scenarios (both in terms of new-physics setup and the parton-level $m^\text{truth}_\ttbar$ cut). This confirms that the global suppression of the efficiency $\varepsilon$ shown in Figure~\ref{fig:quality_notsmeared_1} (relative to $\varepsilon'$) originates solely from the tagging of the hadronic top quark, and is therefore related to the performance of \htt. The latter can thus directly be assessed from the quantity $\varepsilon$, and it is different between SM \ttbar events and those including the interference of top-related SMEFT operators with the SM.

Our results demonstrate that the performance of \htt could be strongly impacted by the physics model that is used as a reference during its tuning. Including effective operators such as those in eq.~\eqref{eq:ops} favours the production of a boosted top-antitop pair more than in the SM, as expected from operators sensitive to the event's energy scale. While in this case the presence of operators not included in the \htt tuning enhances the reconstruction efficiency, this is not generally true, and a tuning based on potential EFT contributions could find different optimal tagging parameters.

Importantly, analyses assuming SM-like \htt reconstruction efficiencies would underestimate the reconstruction and tagging efficiency for any data \ttbar events involving these operators, and would hence systematically overestimate the magnitude of the corresponding Wilson coefficient. This observation reinforces the importance of using operator-dependent reconstruction efficiencies in SMEFT fits to boosted top-quark data.

The presented efficiencies are, however, normalised to the number of events featuring an on-shell \ttbar pair. The obtained increase in the tagging efficiency $\varepsilon$ in the presence of SMEFT operators may, therefore, also be related to a different probability of getting at least one off-shell top quark in the events.
This problem is addressed by the Dalitz-plot heat-maps shown in Figure~\ref{fig:dalitz}, which depict the on-shellness of the produced hadronic top quark. In these figures, we display the correlations between two ratios of invariant masses, $m_{13}/m_{123}$ and $m_{23}/m_{123}$. The three integers 1, 2 and 3 denote the three ($\pt$-ordered) subjets comprised in the hadronically-decaying boosted top quark, so that $m_{123}$ stands for the invariant mass of the three-subjet system, $m_{13}$ for the invariant mass the system made of the leading and third subjets, and $m_{23}$ for that of the system made of the second and third subjets. We present results by restricting the events to those events featuring on-shell top quarks (left column) and for the entire generated samples (right column). Moreover, we explore the difference between the SM (top row), a scenario in which the ${\cal O}_{Qq}^{3,1}$ operator of eq.~\eqref{eq:ops} is turned on (middle row), and a scenario in which the ${\cal O}_{Qq}^{3,8}$ operator of eq.~\eqref{eq:ops} is turned on (bottom row).

As can be seen, the jet combinatorics are correctly resolved in most events in the case of the SM. The leading jet is most often that originating from the two-body $t\to W b$ decay (with the $b$-tagging information being ignored), and the next two jets are those stemming from the hadronic $W$-boson decay. The distribution of the  $m_{23}/m_{123}$ ratio is indeed concentrated around $m_W/m_t$ for the two subfigures of the upper row of Figure~\ref{fig:dalitz}. The spread around this value is more pronounced when no restriction is enforced on the invariant mass of the top quarks at parton level, as observed from a comparison of the predictions shown in the top-left and top-right figures. This can be easily explained by the inefficiency of \htt to correctly tag off-shell top jets, as, by default, the algorithm has been tuned on events featuring on-shell top quarks.

This situation changes slightly when EFT operators are enabled (middle and lower rows of Figure~\ref{fig:dalitz}). First, although the associated amplitude does not feature any intermediate $W$ boson (as the decay of the top quark proceeds via a single four-fermion operator), the interference with the SM diagrams (our predictions being truncated at dimension-six) is sufficient to keep the properties that the leading jet is the $b$-jet, and that the next two jets can be paired to reconstruct a hadronically-decaying $W$ boson. It is additionally noticeable that the effective operators considered affect the reconstructed top quark so that the latter is naturally more often on-shell (and more boosted due to the energy growth inherent to the effective-theory paradigm). Consequently, we can expect better performance of \htt, which confirms what was already found in Figure~\ref{fig:quality_notsmeared_1}.

\subsection{Boosted tops as a probe to new physics in the SMEFT}\label{sec:collider}

\begin{table*}
  \centering
  \renewcommand{\arraystretch}{1.25}
  \sisetup{round-mode=places, round-precision=2}
  \begin{tabular}{@{\extracolsep{4pt}}ll
      @{\quad}S[round-mode=figures,round-precision=3]S[round-mode=places,round-precision=2]
      @{\quad}S[round-mode=figures,round-precision=3]S[round-mode=places,round-precision=2]@{}}
    \toprule
    &	& \multicolumn{2}{c}{Background} &\multicolumn{2}{c}{SM signal} \\
    \cline{3-4} \cline{5-6}\\[-2ex]
    \multicolumn{2}{l}{Selections}
    & \multicolumn{1}{l}{Events} & \multicolumn{1}{c}{$\varepsilon$}
    & \multicolumn{1}{l}{Events} & \multicolumn{1}{c}{$\varepsilon$} \\
    \midrule
    1.&	Initial                        & 124264.50 & \multicolumn{1}{c}{--} & 346061.01 & \multicolumn{1}{c}{--} \\
    2.&	$N_\ell=1$                     & 85506.63 & 0.688 & 227116.93 & 0.656 \\
    3.&	$\pt(\ell_1) > 30$~GeV  & 46668.91 & 0.546 & 133628.45 & 0.588 \\
    4.&	$N_b\geq2$                     & 23515.04 & 0.504 & 85299.59 & 0.638 \\
    5.&	$N_\text{light\ jets}>2$        & 21126.59 & 0.898 & 75787.27 & 0.888 \\
    6.&	$\slashed{E}_\mathrm{T} > 30$~GeV & 18002.61 & 0.852 & 66834.09 & 0.882 \\
    7.&	$N_\text{fat-jet}>0$            & 5231.95 & 0.291 & 53417.98 & 0.799 \\
    8.&	$ N_{t^\text{rec}_H}>0 $ & 274.58 & 0.052 & 24478.32 & 0.458 \\
    9.&	$m^\text{rec}_\ttbar > 950$~GeV     & 164.26 & 0.598 & 23260.59 & 0.950 \\[1ex]
    \midrule
    & $S/B$                         & & & 141.61&\\
    & $S/\sqrt{B}$                  & & & 1814.91&\\
    \bottomrule
  \end{tabular}
  \caption{Number of \ttbar and $Wb\bar b$+jets SM events surviving each step of our analysis, presented together with their respective selection efficiency $\varepsilon$. The results are normalised to an integrated luminosity of 300~fb$^{-1}$. In the last row of the table, we provide two alternative means to assess the analysis significance, namely the $S/B$ and $S/\sqrt{B}$ ratios where $S$ and $B$ are the number of \ttbar and background events passing all cuts.}\label{tab:cutflow}
\end{table*}

In this section, we explore how the findings of Section~\ref{sec:SMEFT_perf} affect the sensitivity of the LHC to SMEFT effects originating from the operators of eq.~\eqref{eq:ops}. We begin by providing, in Table~\ref{tab:cutflow}, the numbers of events surviving each of the selection cuts introduced in the previous section, both for the \ttbar signal and the $W b \bar b$ + jets background. Our results are normalised to an integrated luminosity of \SI{300}{\ifb}, and we additionally estimate the efficiencies associated with each cut, which we define as the ratio of the number of events surviving a given cut to the number of events surviving the previous cut. Whereas the last cut on the invariant mass of the reconstructed \ttbar system (\ie\ the ninth one in the table, $m_{\ttbar}^\text{rec}> \SI{950}{\GeV}$) is not necessary for physics-analysis purposes, it is required to match the Monte Carlo signal-generation cut implemented in Section~\ref{sec:SMEFT_perf} (to enable a more efficient event-generation process in the boosted regime).

As already noticeable from the results introduced earlier in this manuscript, for instance from the invariant-mass spectra displayed in Figure~\ref{fig:leptonic_reco}, the events surviving the entire selection are primarily dominated by signal events, which hence have large expected event-counts. This is further reflected in the $S/B$ and $S/\sqrt{B}$ ratios provided as significance estimators in the lower rows of Table~\ref{tab:cutflow}, these two metrics being evaluated in terms of the number of signal events $S$ and background events $B$ passing all the analysis cuts. The background is thus fully under control in our study, so a shape analysis can be implemented to study how kinematic distributions can be best used to constrain the SMEFT-operators' Wilson coefficients.

To do this, we first increase the final selection cut to maximise sensitivity by probing more deeply boosted top-antitop production. In the following, we hence consider either $m^\text{rec}_{\ttbar} > $ \SI{1}{TeV} or $m^\text{rec}_{\ttbar} >$ \SI{1.5}{TeV}. The sensitivity of the LHC to a given SMEFT operator is derived through the evaluation of a $\chi^2$ test-statistic in an asymptotic scheme that involves deviations of SMEFT predictions relative to the associated SM predictions for a given set of observables. Our analysis explores simultaneously the distributions of the following observables:
\begin{itemize}
\item the invariant mass $m^\text{rec}_\ttbar$ of the di-top system;
\item the transverse momentum $\pt(j^{R=1.5})$ of the leading fat-jet;
\item the transverse momenta $\pt(j^{R=0.4}_1)$, $\pt(j^{R=0.4}_2)$ and $\pt(j^{R=0.4}_3)$ of the three leading small-$R$ jets;
\item the transverse-momentum spectrum $\pt(\tH)$ of the reconstructed hadronic top quark;
\item the transverse-momentum spectrum $\pt(\tL)$ of the reconstructed leptonic top quark;
\item the rapidity difference $\Delta y (\tL,\tH)$ between the two reconstructed top quarks;
\item and the azimuthal-angle difference $\Delta\varphi(\tL,\tH)$ between the two reconstructed top quarks.
\end{itemize}

In order to estimate the $\chi^2$ value associated with a specific SMEFT scenario, each of the nine histograms considered was divided into 25~bins (20~and 16~for the $\Delta y (\tL,\tH)$ and $\Delta\varphi(\tL,\tH)$ distributions respectively), and we calculated the quantity
\begin{equation}\label{eq:chi2}
  \chi^2 = \sum_{i} \frac{\left( N^\mathrm{exp}_i -N^\mathrm{obs}_i\right)^2}{\sqrt{N^\mathrm{obs}_i + \left(\Delta_\mathrm{sys} N^\mathrm{obs}_i \right)^2}}\,,
\end{equation}
in which we sum over all bins and all histograms. The SM predictions are taken as the null hypothesis, $N^\mathrm{exp}_i$ denoting hence the expected number of events in the SM for a given observable and bin $i$, $N^\mathrm{obs}_i$ standing for the corresponding SMEFT predictions, and $\Delta_\mathrm{sys} N^\mathrm{obs}_i$ referring to the error on the SMEFT predictions. In other words, we enforce that the pseudo-data corresponding to the SM scenario (\ie~the origin of the Wilson coefficient parameter space) corresponds to the background expectation with suppressed statistical and systematical fluctuations, which consists, therefore, of an Asimov dataset. The above $\chi^2$ test is thus asymptotically equivalent to a profile likelihood ratio $\Delta{\chi^2} = \chi^2_{\rm SMEFT} - \chi^2_\text{best}$ for a given SMEFT scenario with an implicit best-fit reference model evaluated in the case of the SM (therefore with $\chi^2_\text{best} = 0$). Without explicitly performing any profiling, we thus estimate the sensitivity of a profile-likelihood fit by comparing the obtained $\chi^2$ values with that expected from a $\chi^2$ distribution with one degree of freedom. In practice, however, profiled constraints could be slightly weaker due to a less perfect fit of observed data to the background model.

\begin{table*}
  \renewcommand{\arraystretch}{1.35}
  \centering
    \begin{tabular}{@{\extracolsep{4pt}}lc@{\hspace{2em}}c@{\hspace{2em}}c@{\hspace{2em}}c@{\hspace{2em}}}
      \toprule
      Mass req. & \multicolumn{4}{c}{$m^\text{rec}_\ttbar > \SI{1}{\TeV}$}\\
      \cline{2-5}\\[-2ex]
      Lumi. [\si{\iab}] & \multicolumn{2}{c}{$0.3$} & \multicolumn{2}{c}{$3$} \\
      \cline{2-3} \cline{4-5}\\[-2ex]
      $ \Delta_\mathrm{\,sys} $ [\%] & 0 & 10 & 0 & 10\\
      \midrule
      $ \mathcal{O}_{Qq}^{1,8} $ & $ \pt(\tL) $ & $\pt(\tL)~~~~$ & $\pt(\tL)$ & $\pt(\tL)$\\
      $ \mathcal{O}_{Qq}^{3,8} $ & $ \pt(\tL) $ & $\pt(\tL)$ & $\pt(\tL)$ & $\pt(\tL)$\\
      $ \mathcal{O}_{Qq}^{3,1} $ & $ \pt(\tL) $ & $\pt(\tL)$ & $\pt(\tL)$ & $\pt(\tL)$\\
      $ \mathcal{O}_{tu}^{8}  $ & $ \pt(\ell_1) $ & $\pt(\ell_1)$ & $\pt(\ell_1)$ & $\pt(\ell_1)$\\
      $ \mathcal{O}_{td}^{8}  $ & $ \pt(\ell_1) $ & $\pt(\ell_1)$ & $\pt(\ell_1)$ & $\pt(\ell_1)$\\
      $ \mathcal{O}_{Qu}^{8}  $ & $ \pt(\tL) $ & $\pt(\tL)$ & $\pt(\tL)$ & $\pt(\tL)$\\
      $ \mathcal{O}_{Qd}^{8}  $ & $ \pt(\tL) $ & $\pt(\tL)$ & $\pt(\tL)$ & $\pt(\tL)$\\
      $ \mathcal{O}_{tq}^{8}  $ & $ \pt(\ell_1) $ & $\pt(\ell_1)$ & $\pt(\ell_1)$ & $\pt(\ell_1)$\\[1ex]
      \bottomrule
    \end{tabular}

    \vspace*{.3cm}
    \begin{tabular}{@{\extracolsep{4pt}}lc@{\hspace{2em}}c@{\hspace{2em}}c@{\hspace{2em}}c@{\hspace{2em}}}
      \toprule
      Mass req. & \multicolumn{4}{c}{$m^\text{rec}_\ttbar > \SI{1.5}{\TeV}$}\\
      \cline{2-5}\\[-2ex]
      Lumi. [\si{\iab}] & \multicolumn{2}{c}{$0.3$} & \multicolumn{2}{c}{$3$} \\
      \cline{2-3} \cline{4-5}\\[-2ex]
      $ \Delta_\mathrm{\,sys} $ [\%] & 0 & 10 & 0 & 10\\
      \midrule
      $ \mathcal{O}_{Qq}^{1,8} $ & $ \pt(\tH) $ & $\pt(\tL)$ & $ \pt(\tH) $ & $ \Delta y_\ttbar $\\
      $ \mathcal{O}_{Qq}^{3,8} $ & $ \pt(\tH) $ & $ \pt(j^{R=1.5}_1) $ & $ \pt(\tH) $ & $\pt(\tL)$\\
      $ \mathcal{O}_{Qq}^{3,1} $ & $\pt(\tL)$ & $\pt(\tL)$ & $\pt(\tL)$ & $ \Delta y_\ttbar $\\
      $ \mathcal{O}_{tu}^{8}  $ & $ \pt(\ell_1) $ & $\pt(\ell_1)$ & $\pt(\ell_1)$ & $\pt(\ell_1)$\\
      $ \mathcal{O}_{td}^{8}  $ & $ \pt(\ell_1) $ & $\pt(\ell_1)$ & $\pt(\ell_1)$ & $\pt(\ell_1)$\\
      $ \mathcal{O}_{Qu}^{8}  $ & $\pt(\tL)$ & $\pt(\tL)$ & $\pt(\tL)$ & $\pt(\tL)$\\
      $ \mathcal{O}_{Qd}^{8}  $ & $ \pt(\tH) $ & $ \pt(\tL) $ & $ \pt(\tH) $ & $ \Delta y_\ttbar$ \\
      $ \mathcal{O}_{tq}^{8}  $ & $ \pt(\ell_1) $ & $\pt(\ell_1)$ & $\pt(\ell_1)$ & $\pt(\ell_1)$\\[1ex]
      \bottomrule
    \end{tabular}
  \caption{Observable driving the sensitivity of the LHC (at 68\% confidence level) to a given SMEFT operator from eq.~\eqref{eq:ops} (first column). We consider both a perfect situation without systematics ($\Delta_\mathrm{\,sys} = 0$, second and fourth columns), and one with 10\% of systematics ($\Delta_\mathrm{\,sys}=10\%$, third and fifth columns). Moreover, we present results for \SI{300}{\ifb} and \SI{3}{\iab}, and for an invariant mass cut of $m^\text{rec}_\ttbar > \SI{1}{\TeV}$ (upper panel) and $m^\text{rec}_\ttbar > \SI{1.5}{\TeV}$ (lower panel). \label{tab:best_histo}}
\end{table*}

In Table~\ref{tab:best_histo}, we provide information on the observable found to provide the strongest sensitivity to each SMEFT operator. The results are shown for the two cuts on the invariant mass considered, $m^\text{rec}_\ttbar > \SI{1}{\TeV}$ (upper panel of Table~\ref{tab:best_histo}) and $m^\text{rec}_\ttbar > \SI{1.5}{\TeV}$ (lower panel of Table~\ref{tab:best_histo}). Moreover, we consider LHC luminosities of \SI{300}{\ifb} and \SI{3000}{\ifb}, and two different options for the amount of systematics $\Delta_\mathrm{\,sys}$ used in eq.~\eqref{eq:chi2}. We take as a reference the ideal situation in which there are no systematic uncertainties ($\Delta_\mathrm{\,sys} =0$), as well as a more realistic situation in which we set $\Delta_\mathrm{\,sys} = 10\%$. In our procedure to extract this information, we define the sensitivity on the basis of a 68\% confidence level. When we consider a moderate definition of the boosted regime with $m^\text{rec}_\ttbar > \SI{1}{\TeV}$, the sensitivity is always driven by the distribution in the transverse momentum of either the leptonically-decaying top quark ($\pt(\tL)$) or of the lepton originating from the decay of this top quark ($\pt(\ell_1)$). The information brought by the hadronic branch of the event is found to be sub-leading for all SMEFT operators and systematic-uncertainty assumptions. However, the situation changes when the boosted regime is probed more deeply through the tighter cut $m^\text{rec}_\ttbar > \SI{1.5}{\TeV}$. Here, both top quarks are reconstructed and tagged more accurately (in particular through the better performance of \htt in a SMEFT scenario, see Section~\ref{sec:SMEFT_perf}). This leads to an increased discovery potential through use of a larger set of contributing observables. This statement is illustrated in the lower panel of the table, which displays a greater variability in the leading observable driving the sensitivity of the LHC to a given SMEFT operator, with the $\mathcal{O}_{Qq}^{1,8}$, $\mathcal{O}_{Qq}^{3,8}$, $\mathcal{O}_{Qq}^{3,1}$, and $\mathcal{O}_{Qd}^{8}$ operators now most sensitive to either hadronic-top or \ttbar-system observables.

\begin{figure*}
  \centering
  \includegraphics[width=0.48\textwidth]{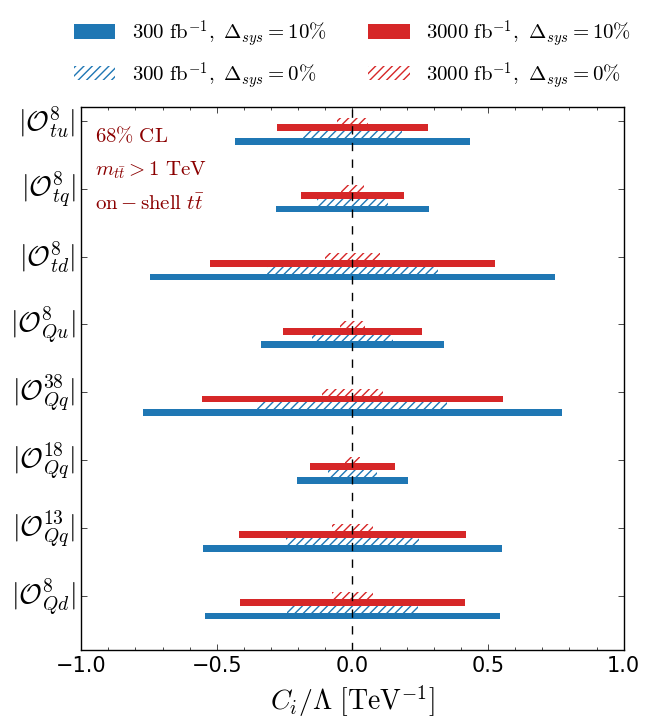}
  \includegraphics[width=0.48\textwidth]{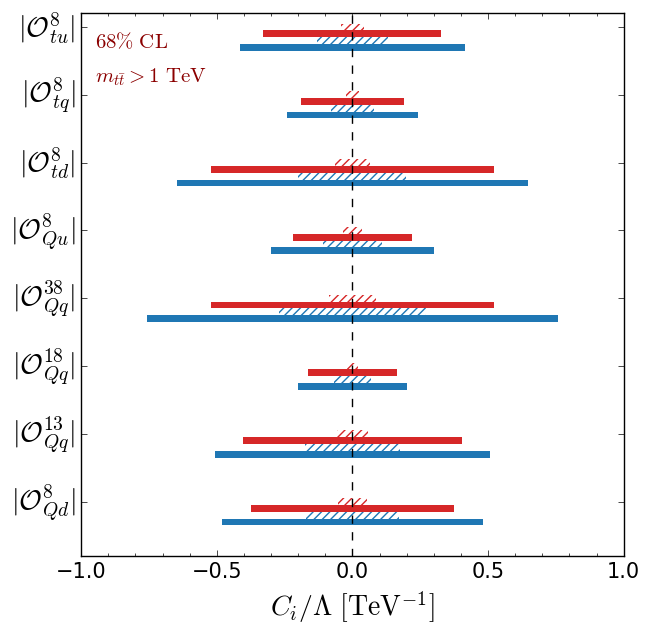}\\
  \includegraphics[width=0.48\textwidth]{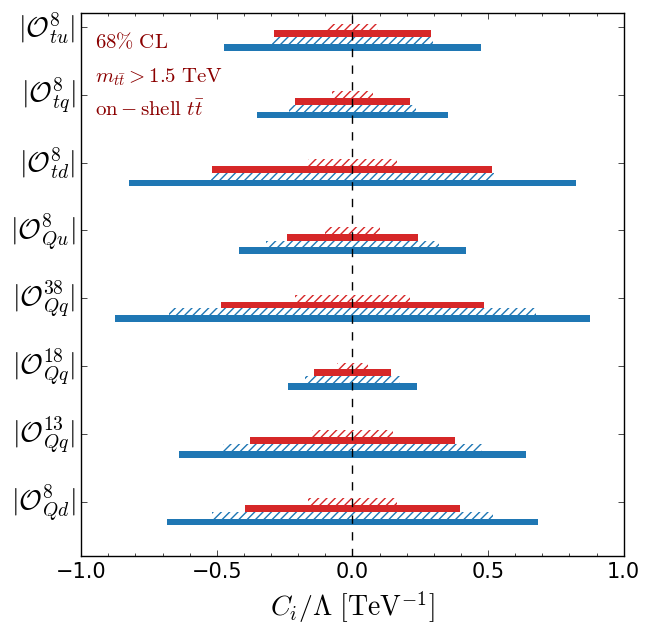}
  \includegraphics[width=0.48\textwidth]{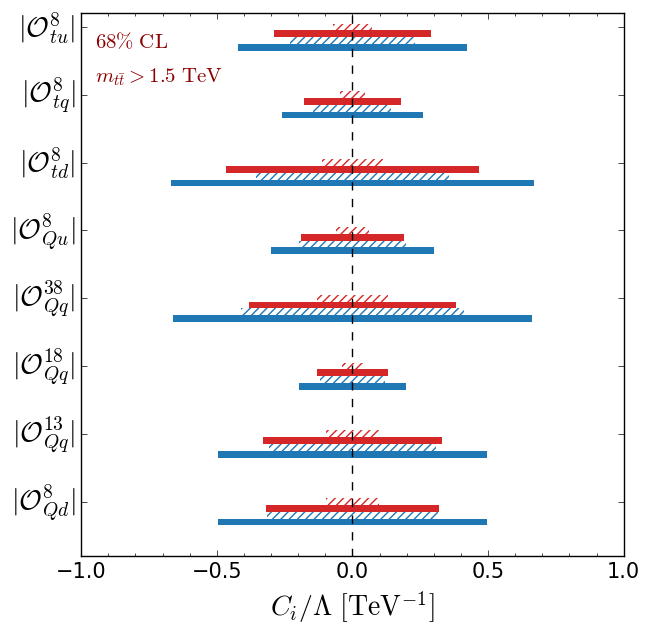}\\
  \caption{Sensitivity of the LHC to the various SMEFT operators of eq.~\eqref{eq:ops}. We present predictions for \SI{300}{\ifb} (blue) and \SI{3000}{\ifb} (red), $\Delta_\mathrm{\,sys} = 0$ (shaded bars) and 10\% (solid bars), and we distinguish an analysis of the full \ttbar event sample generated (right column) and after enforcing on-shell top-antitop production (left column). Two analysis cuts on the invariant mass of the reconstructed top pair are imposed, $m^\text{rec}_\ttbar > \SI{1}{\TeV}$ (upper panel) and $m^\text{rec}_\ttbar > \SI{1.5}{\TeV}$ (lower panel).\label{fig:sensitivity}}
\end{figure*}

Our final projections of SMEFT Wilson-coefficient expected limits, assuming the SM, are shown in Figure~\ref{fig:sensitivity}. We derive the sensitivity of the LHC to each of the operators considered, making use of the procedure described above. We present bounds on the associated Wilson coefficients, both for an integrated luminosity of \SI{300}{\ifb}  (blue) and \SI{3000}{\ifb} (red), and for the two options explored for the level of systematics, namely $\Delta_\mathrm{\,sys} = 0$ (shaded bars) and 10\% (solid bars). In addition, we distinguish the case in which we pre-select at parton-level on-shell \ttbar events (left subfigures) and that in which we analyse the full event sample generated (right subfigures). As for the previous discussion, we first implement a relatively inclusive requirement of \SI{1}{\TeV} on the invariant mass of the reconstructed \ttbar system (upper row) and as well as a more stringent $m^\text{rec}_\ttbar > \SI{1.5}{\TeV}$ cut (bottom row).

We find limits on  $|C/\Lambda|$ that lie in the 0.1--\SI{1}{\per\TeV} range. This means that for Wilson coefficients satisfying $C\sim 1$, effective scales in the 1--5~TeV range can be probed. Conversely, for TeV-scale new physics, couplings of $\mathcal{O}(0.1)$ can be reached. The bounds are found to be mildly more constraining with the increase in luminosity as well as with a harder cut on $m^\text{rec}_\ttbar$, as expected, and the impact of off-shell top-antitop production is additionally found to be sub-leading. Such a sensitivity is of comparable size with that estimated on the basis of global fits (see \eg~predictions from Ref.~\cite{Brivio:2019ius}), which demonstrates the potential of including dedicated analyses of boosted top-quark pair production and decay in SMEFT global fits. Global fits of LHC Run 2 data indeed indicate that $|C/\Lambda|$ has to be smaller than about $0.1$--\SI{1}{\per\TeV} too. Our results should however additionally be compared with individual limits extracted from fits of a large set of observables when one SMEFT operator is considered at a time (for a fairer comparison). Such fits lead to bounds on $|C/\Lambda|$ of $\mathcal{O}(0.1)\SI{}{\per\TeV}$~\cite{Ethier:2021bye}, which are thus comparable with the findings of Figure~\ref{fig:sensitivity}. Whereas exploiting boosted top quark production is already known to have a strong constraining power on individual operators (for instance in the context of top dipole moments, where it has been shown to significantly improve the bounds by a factor of a few~\cite{Aguilar-Saavedra:2014iga}), a detailed quantitative analysis of its impact lies beyond the scope of this paper. Here, we have only investigated how using a specific boosted-top quark channel could lead to a better assessment of the sensitivity of the LHC to top-quark-related SMEFT operators, thanks to a joint usage of a variety of potentially relevant observables and improved top-tagging capabilities in the SMEFT.

\section{Conclusion and outlook}\label{sec:conclusion}
Jet substructure methods are known to be among the key players in the search for new phenomena beyond the Standard Model of particle physics. Among these, a set of dedicated techniques are related to the identification of jets originating from the hadronic decay of a boosted top quark. In this paper, we have reported the development of an interface between the \htt package and two software tools widely used in the high-energy physics community, namely the \ma and \rivet frameworks. Thanks to this development, the many users of these platforms now have the possibility to exploit boosted hadronically-decaying top quarks and their properties in analyses of high-energy physics events for the Large Hadron Collider and beyond.

We have briefly described these two implementations and how to use them. Our developments equip the \rivet toolkit from version~3.1.7, which is available from \textsc{HepForge} (see \url{https://rivet.hepforge.org/}), as well as the \ma framework from version~2.0.4, available from \textsc{GitHub} (see \url{https://github.com/MadAnalysis/madanalysis5/releases}). Moreover, detailed tutorials exploiting all the possibilities can be found in the \lstinline{"analyses/examples/EXAMPLE_HTT.cc"} analysis file shipped with \rivet, as well as in the \ma tutorial available from \url{https://github.com/MadAnalysis/tutorial_osu}.

To illustrate the power of these developments, we have considered the SMEFT framework in which new physics manifests through non-renormalisable operators in the Standard Model fields. We have focused on eight dimension-six, four-fermion operators relevant to the top-quark sector, chosen as they are not stringently constrained by current SMEFT global fits. The analysis of the production of pairs of boosted top quarks could therefore provide new handles on associated heavy BSM physics. We have explored this option by first investigating the performance of the \htt algorithm in the presence of non-vanishing SMEFT operators. Whereas the algorithm is tuned on SM top-pair production and decay, we have observed that its performance improves further in the presence of the considered additional SMEFT operators in the model's Lagrangian. The energy dependence of the SMEFT operators considered indeed favours the production of very energetic boosted top quarks, with properties enhancing their tagging possibility by the \htt method. This observation highlights the importance of considering new-physics effects upon reconstruction performance when attempting SMEFT parameter fits.

Secondly, we have investigated differential observables in boosted top-antitop production following \htt tagging, to study how deviations from the Standard Model can best be used to isolate SMEFT effects emerging from the new operators. We have shown that a simple analysis based on \htt could lead to bounds comparable with those stemming from other means to constrain SMEFT operators. We hope that this demonstrates the potential of the developments presented in this work and that they will serve the community well in the future.

\section*{Acknowledgements}
This work has been partly supported by the French ANR (grant ANR-21-CE31-0013,
`DMwithLLPatLHC'), by the UK Royal Society (grant UF160548) and STFC (grant
ST/S000887/1), and by a short-term studentship funded by the European Union's
Horizon 2020 research and innovation programme as part of the Marie
Sklodowska-Curie Innovative Training Network MCnetITN3 (grant agreement
no.~722104).

\bibliographystyle{JHEP}
\bibliography{bibliography}

\end{document}